\documentclass[fleqn,11pt]{wlscirep}
\usepackage[utf8]{inputenc}
\usepackage[T1]{fontenc}

\usepackage{caption}
\usepackage{subcaption}
\usepackage{nameref}
\usepackage{csquotes}

\newcommand{\C}[1]{{\color{black!40!black} {#1}}} 
\newcommand{\E}[1]{{\color{black!40!black} {#1}}} 

\newcommand{\CC}[1]{{\color{black!40!black} {#1}}}  
\newcommand{\EE}[1]{{\color{black!40!black} {#1}}} 

\usepackage[style=numeric-comp, sorting=none]{biblatex}
\usepackage{eucal}

\newcommand{\beginsupplement}{%
        \setcounter{table}{0}
        \renewcommand{\tablename}{\textbf{Supplementary Table}}
        \renewcommand{\thetable}{\arabic{table}}%
        \setcounter{figure}{0}
        \renewcommand{\figurename}{\textbf{Supplementary Figure}}
        \renewcommand{\thefigure}{\arabic{figure}}%
     }

\makeatother

\title{\EE{Learning to Simulate High Energy Particle Collisions from Unlabeled Data}}

\author[1,*]{Jessica N. Howard}
\author[2]{Stephan Mandt}
\author[1]{Daniel Whiteson}
\author[2]{Yibo Yang}
\affil[1]{Department of Physics \& Astronomy, UC Irvine, Irvine, CA\E{, USA.}}
\affil[2]{Department of Computer Science, UC Irvine, Irvine, CA\E{, USA.}}

\affil[*]{jnhoward@uci.edu}
\affil[ ]{ }
\affil[ ]{Authors are listed alphabetically.}

\begin{abstract}
\CC{In many scientific fields which rely on statistical inference, simulations are often used to map from theoretical models to experimental data, allowing scientists to test model predictions against experimental results. }
Experimental data is often reconstructed from indirect measurements causing the aggregate transformation from theoretical models to experimental data to be poorly-described analytically. 
Instead, numerical simulations are used at great computational cost.
\E{We introduce Optimal-Transport-based Unfolding and Simulation (OTUS), a fast simulator based on unsupervised machine-learning that is capable of predicting experimental data from theoretical models.} 
\C{Without the aid of current simulation information, OTUS trains a probabilistic autoencoder to transform directly between theoretical models and experimental data.}
Identifying the probabilistic autoencoder's latent space with the space of theoretical models causes the decoder network to become a fast, predictive simulator with the potential to replace current, computationally-costly simulators. 
\E{Here, we provide proof-of-principle results on two particle physics examples, $Z$-boson and top-quark decays, but stress that OTUS can be widely applied to other fields.}
\end{abstract}

\begin{document}

\flushbottom
\maketitle
\thispagestyle{empty}

\section{Introduction} \label{sec:intro}
\CC{
From measuring masses of particles to deducing the likelihood of life elsewhere in the Universe, a common goal in analyzing scientific data is statistical inference --- drawing conclusions about values of a theoretical model's parameters, $\theta$, given observed data, $x$. The likelihood model of observed data, $p(x|\theta)$, is a central ingredient in both frequentist and Bayesian approaches to statistical inference; however, it is typically intractable, due to the complexity of a full probabilistic description of the data generation process. One way to circumvent this difficulty is to \emph{simulate} experimental data for a given value of the theoretical parameters, $\theta$, from which a probability model of the likelihood, $p(x|\theta)$, can be constructed and used for downstream statistical inference regarding $\theta$. This is known as simulation-based inference, and has found application across scientific disciplines ranging from particle physics to cosmology~\cite{cranmer2020frontier}.
}

\CC{
However, traditional approaches to simulation, which attempt to faithfully model complex physical phenomena, can be computationally expensive --- a limitation we aim to overcome in this work. In simulation-based inference, experimental data arising from a physical system typically depend on an initial configuration of the system, $z$, that is unobserved, or belonging to a \emph{latent} space, while the parameters $\theta$ govern the underlying mechanistic model.
In many cases, the transformation from the latent state to experimental data is non-trivial, involving complex physical interactions that cannot be described analytically, but can be \emph{simulated} numerically by Monte-Carlo algorithms.  In particle physics, for example, the parameters $\theta$ govern theoretical models that describe fundamental particle interactions. These fundamental interactions produce secondary particles, $z$, which are not directly observable and often transform in flight before passing through layers of detectors whose indirect measurements, $x$, can help reconstruct their identities and momenta. The transformation from the unobserved latent space, particles produced in the initial interaction, to the experimental data, is stochastic, governed by quantum mechanical randomness, and has no analytical description.
}

Instead, Monte-Carlo-based numerical simulations of in-flight and detection processes generate samples of possible experimental data for a given latent \C{space configuration}~\C{\cite{geant, pythia, delphes, mcexpenseATLAS, mcexpenseCMS}}. This approach is computationally expensive~\cite{mcexpenseATLAS, mcexpenseCMS} because it requires the propagation and simulation of every individual particle, each creating subsequent showers of thousands of derivative particles. Additionally, these simulations contain hundreds of parameters which must be \C{extemporaneously tuned} to give reasonable results in \E{control regions} of the data where the latent space has been well-established by results from previous experiments.

In particle physics, like many other fields in the physical sciences~\cite{otherfields_chem, otherfields_climate, otherfields_cosmology, otherfields_photonics}, the computational cost of numerical simulations has become a central bottleneck. A fast, interpretable, flexible, data-driven generative model which can transform between the latent space and the experimental data would be significant for these fields. Recent advances in the flexibility and capability of machine learning (ML) models have allowed for their application as computationally inexpensive \C{simulators~\cite{calogan, ganlhc, DIBGANVAE, lagan, highVAE, e2eSAE, Hashemi2019, SARM, Andreassen_2019}}. Applications of these techniques have made progress towards this goal but fall short in crucial ways. For example, approaches leveraging Generative Adversarial Networks (GANs) are able to mimic experimental data for fixed distributions in the latent space~\cite{calogan, ganlhc, DIBGANVAE}, but are unable to generate predictions for new values of latent variables, a crucial requirement for a simulator. Other efforts condition on latent variables~\cite{lagan} but require training with labeled pairs generated by slow Monte-Carlo generators, incurring \C{some of} the computational cost they seek to avoid.
    
\begin{figure}[h!t]
    \centering
        \vspace{1cm}
    \includegraphics[width=0.6\textwidth]{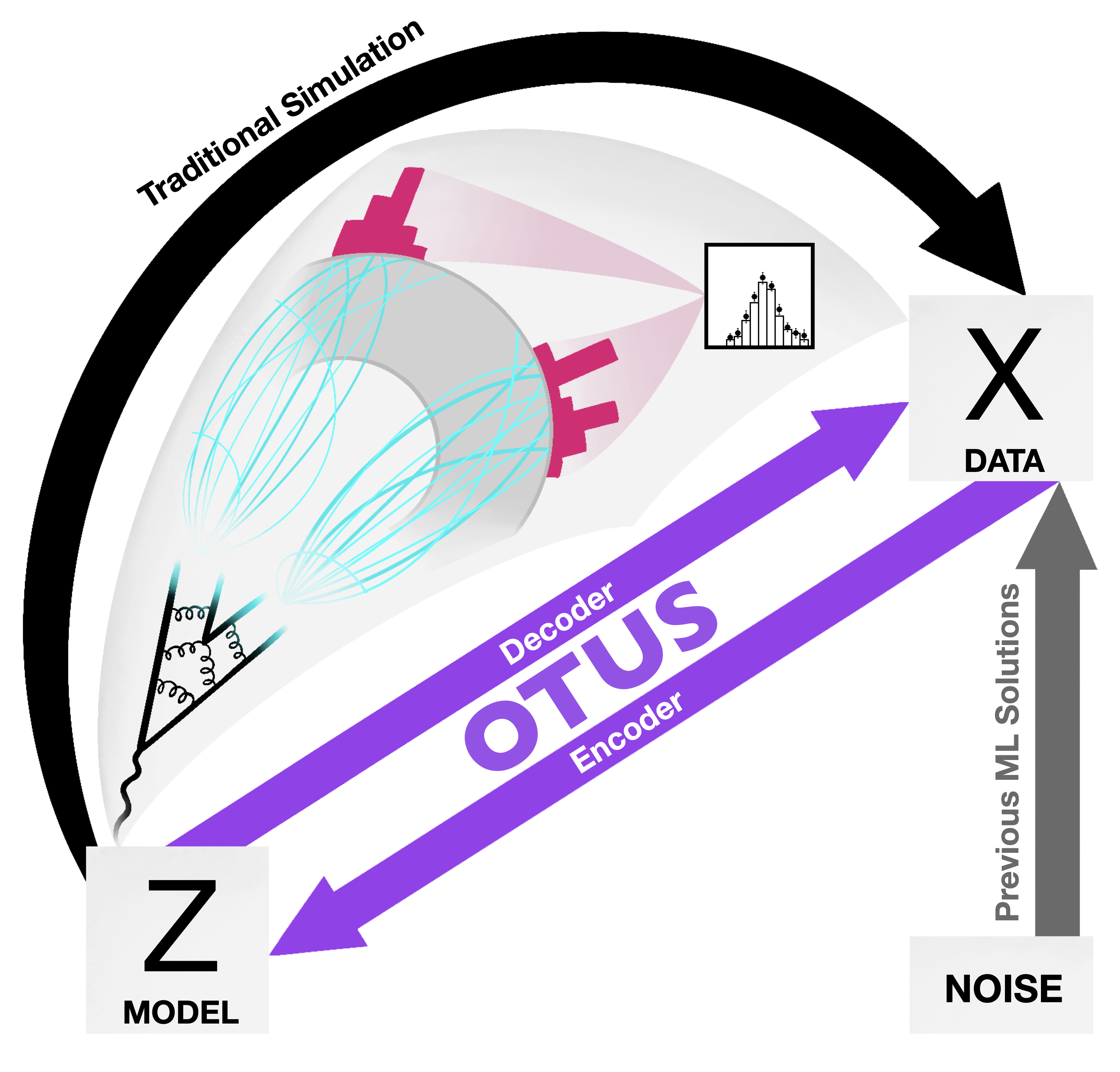}
    \caption{\E{\textbf{Schematic of the problem and the solution.}} Current simulations map from a physical latent space, $\mathcal{Z}$, to data space, $\mathcal{X}$, attempting to mimic the real physical processes at every step. This results in a computationally intensive simulation. Previous \E{Machine Learning (ML)} solutions can reproduce the distributions in $\mathcal{X}$ but are not conditioned on the information in $\mathcal{Z}$\C{; instead they map from unphysical noise to $\mathcal{X}$,} which limits their scope. We introduce a new method which provides the best of both worlds. \C{OTUS provides a simulation $\mathcal{Z} \rightarrow \mathcal{X}$ (Decoder) which is conditioned on $\mathcal{Z}$ yet is computationally efficient. Advantageously, it also inadvertently provides an equivalently fast unfolding mapping from $\mathcal{X} \rightarrow \mathcal{Z}$ (Encoder).} 
    }
    \label{fig:schema}
\end{figure}

We lay the foundations and provide a proof-of-principle demonstration \E{for Optimal-Transport-based Unfolding and Simulation (OTUS)}. We use unsupervised learning to build a flexible description of the transformation from latent space, $\mathcal{Z}$, to experimental data space, $\mathcal{X}$, relying on theoretical priors, $p(z)$, where $z \in \mathcal{Z}$ and \C{a set of samples of experimental data $\{x \in \mathcal{X}\}$} but, crucially, no labeled pairs, \C{$(z,x)$}.   Our model applies \E{a type} of probabilistic autoencoder~\cite{swae, wae}, which \E{learns} two mappings: encoder (data $\rightarrow $ latent, $p_E(z \mid x)$) and decoder (latent $\rightarrow $ data, $p_D(x \mid z)$). Typical probabilistic autoencoders (i.e. \E{variational} autoencoders (VAEs)~\cite{vae, DIBGANVAE}) use a \E{ simple}, \E{ unphysical} latent space, $\mathcal{Y}$, for computational tractability during learning. However, this causes VAEs to suffer from the same weakness as GANs: doomed to mimic the data distribution, $p(x)$, for a fixed physical latent space, $p(z)$, unless the model compromises to requiring expensive simulated pairs (e.g. a conditional VAE approach~\cite{conditionalVAE}). OTUS's innovation is to align the probabilistic autoencoder's latent space, $\mathcal{Y}$, with that of our inference task, $\mathcal{Z}$. With this change, our decoder becomes a computationally inexpensive, conditional simulator mapping $\mathcal{Z} \rightarrow \mathcal{X}$ as well as a tractable transfer function, $p_D(x \mid z)$. \EE{See Fig.~\ref{fig:schema} for a visual description.}
    
For VAEs, identifying $\mathcal{Y}$ with $\mathcal{Z}$ is difficult because the training objective requires the ability to explicitly compute the latent space prior, $p(y)$ for $y \in \mathcal{Y}$. In particle physics, such explicit computations are intractable. We therefore turn to a new form of probabilistic autoencoder: the Sliced Wasserstein Autoencoder (SWAE)~\cite{swae, wae}, which alleviates this, and other, issues by reformulating the objective using the Sliced Wasserstein distance and other ideas from optimal transport theory. This reformulation lets us identify $\mathcal{Y}$ with $\mathcal{Z}$ and also allows the encoder and decoder network mappings to be inherently stochastic.
    
We suggest that an SWAE~\cite{swae} can be used to achieve the broad goal of simulators: learning the mapping from the physical latent space to experimental data directly from samples of experimental data \C{$\{x \sim p(x)\}$} and theoretical priors \C{$\{ z \sim p(z)\}$} in control regions. The resulting decoder ($\mathcal{Z} \rightarrow \mathcal{X}$) can be applied as a simulator, generating samples of experimental data from latent variables in a fraction of the time, and \E{probed} and visualized to ensure a physically meaningful transformation.
Additionally, the decoder's numerically tractable detector response function, $p_D(x \mid z)$, would be useful in other applications, such as direct calculation of likelihood ratios via integration~\cite{Abazov_2004}. The encoder network's $\mathcal{X} \rightarrow \mathcal{Z}$ mapping can also be used in unfolding studies~\cite{unfold_INN, unfold_OmniFold}. Lastly, the mathematical attributes of the SW distance allow for \E{the} inclusion of informed constraints on the mappings.

\CC{In this work, we first present background on the problem, the objective, and discuss related work. In~\nameref{sec:solution}, we present the foundations for the OTUS method and discuss steps toward scaling OTUS to a full simulation capable of replacing current Monte-Carlo methods in particle physics analyses. In~\nameref{sec:results_}, we give initial proof-of-principle demonstrations on $Z$-boson and semileptonic top-quark decays. In~\nameref{sec:methods}, we discuss the details of our methods. We then conclude by discussing directions for future work and also briefly discuss how OTUS might be applied to problems in other scientific fields.}

\section{\CC{Theoretical Background}} \label{sec:simulations} 

The primary statistical task in particle physics, as in many areas of science, is inferring \C{the value of a model parameter, $\theta$, based on a set of experimental data, $\{x\}$.} For example, physicists inferred the mass of the Higgs boson from Large Hadron Collider \C{data~\cite{higgs_ATLAS, higgs_CMS}}.  Inference about $\theta$ requires a statistical model, $p(x \mid \theta)$, \C{which can be used to calculate the probability to make an observation, $x$, given a parameter value, $\theta$.}  Unfortunately, such analytical expressions are unavailable due to the indirect nature of observations and the complexity of detectors. Previous solutions to this problem have relied on numerical Monte-Carlo-based simulations~\cite{geant, pythia, delphes}.

Fundamental particle interactions, like the decay of a Higgs boson, produce a set of particles which define an unobserved latent space, $\mathcal{Z}$. The statistical model $p(z \mid \theta)$ is usually well-understood and can often be expressed analytically or approximated numerically. However, experimenters only have access to \C{samples of experimental data, $\{x\}$}. Therefore, calculating $p(x \mid \theta)$ requires integrating over the unobserved \C{$\{z \sim p(z \mid \theta)\}$}; namely, $p(x \mid \theta) = \int dz\ p(x \mid z)\ p(z \mid \theta)$. 

The transfer function, $p(x \mid z)$, represents the multi-staged transformation from the unobserved latent space, $\mathcal{Z}$, to the experimental data space, $\mathcal{X}$. As latent space particles travel they may \C{decay, interact, or radiate} to produce subsequent showers of hundreds of secondary particles. These particles then pass through the detector, comprising many layers and millions of sensors resulting in a high-dimensional response of order $\mathcal{O}(10^8)$. Finally, the full set of detector measurements are used to reconstruct an estimate of the identities and momenta of the original unobserved particles in the latent space. \C{For the vast majority of analyses this final}, experimental data space, $\mathcal{X}$, has a similar dimensionality%
\footnote{The dimensionality is not necessarily equal due to the imperfect nature of the detection process. For example, $\mathcal{Z}$ may represent four quarks but $\mathcal{X}$ may only contain three jets.} to that of $\mathcal{Z}$, usually $\mathcal{O}(10^1)$. However, the complex, stochastic, and high-dimensional nature of the transformation makes it practically impossible to construct a closed-form expression for the transfer function $p(x \mid z)$. Instead, particle physicists use simulations as a proxy for the true transfer function.

To arrive at $p(x \mid \theta)$, samples of \C{$\{z \sim p(z \mid \theta)\}$} are transformed via simulations into effective samples of \C{$\{x \sim p(x \mid \theta)\}$}, approximating the integral above. Current state-of-the-art simulations strive to faithfully model the details of particle propagation and decay via Monte-Carlo techniques. This approach is computationally expensive and limited by our poor understanding of the processes involved.  Ad-hoc parameterizations often fill gaps in our knowledge but introduce arbitrary parameters which must be tuned to give realistic results using data from \E{control regions}, where the underlying $p(z \mid \theta)$ is well-established from previous experiments, freeing $p(x \mid \theta)$ of surprises. Examples of control regions include decays of heavy bosons (e.g. $Z$) or the top quark ($t$).

The computational cost of current simulations is the dominant source of systematic uncertainties and the largest bottleneck in testing new models of particle physics~\cite{syserr}. A computationally-inexpensive, flexible simulator which can map from $\mathcal{Z}$ to $\mathcal{X}$ such that it effectively approximates $p(x \mid z)$ would be a \E{breakthrough}. 

\section{\CC{Objective and Related Work}} \label{sec:objective}

The development of OTUS was guided by the goals of the simulation task and the information available for training. Specifically, the simulator has access to samples from model priors\E{,} $p(z \mid \theta)_{\textrm{~control}}$\E{,} and experimental data samples\E{,} \C{$\{x_{\textrm{~control}}\}$}. Critically, \C{$\{x_{\textrm{~control}}\}$} samples come from experiments, where the true \C{$\{z_{\textrm{~control}}\}$} are unknown, such that no $(z_{\textrm{~control}},x_{\textrm{~control}})$  pairs exist. Instead, the distribution of \C{$\{z_{\textrm{~control}}\}$} are known to follow $p(z \mid \theta)_{\textrm{~control}}$ and the distribution of \C{$\{x_{\textrm{~control}}\}$} is observed.

The simulator should learn a stochastic transformation $\mathcal{Z} \rightarrow \mathcal{X}$ such that samples \C{$\{z\}$} drawn from $p(z \mid \theta)_{\textrm{~control}}$ can be transformed into samples \C{$\{x\}$} whose distribution matches that of the experimental data \C{$\{x_{\textrm{~control}}\}$}. Additionally, these control regions should be robust so that the simulator can approximate $p(x \mid \theta)$ for different, but related, values of $\theta$. Traditional Monte-Carlo simulators such as GEANT4~\cite{geant} face \C{related challenges}.

The flexibility of ML models at learning difficult functions across a wide array of contexts suggests that these tools could be used to develop a fast simulator.  The objectives described above translate to four constraints on the class of ML model and methods of learning.  Generating samples of \C{$\{x \in \mathcal{X}\}$} requires a (1) \E{generative} ML method. For $z \in \mathcal{Z}$, the simulator maps $z \rightarrow x$ such that the output $x$ depends on the input $z$, meaning the mapping is (2) \E{conditional}. 
The problem's inherent and unknown randomness prevents us from assuming any particular density model, suggesting that our simulator should preferably be (3) \E{inherently stochastic}.
The lack of \C{$(z,x)$}  pairs mandates an (4) \E{unsupervised} training scheme. Additionally, the chosen method should produce a simulation mapping ($\mathcal{Z} \rightarrow \mathcal{X}$) which is inspectable and physically interpretable.

Generative ML models can produce realistic samples of  data in many settings, including natural images. Generative Adversarial Networks (GANs) transform noise into artificial data samples and have been adapted to particle physics simulation tasks for both high-level and raw detector data, which can resemble images~\cite{calogan,  ganlhc, DIBGANVAE, lagan, Hashemi2019}. 
However, while GANs have successfully mimicked existing datasets, \C{$\{x\}$}, for a fixed set of \C{$\{z\}$}, they have not learned the general transformation $z \rightarrow x$ prescribed by $p(x \mid z)$, and so cannot generate fresh samples \C{$\{x'\}$} for a new set of \C{$\{z'\}$}, thus failing condition (2).  Other GAN-based approaches~\cite{lagan} condition the generation of \C{$\{x\}$} on values of \C{$\{z\}$}, but in the process use labeled pairs \C{$(x, z)$}, which are only obtained from other simulators, rather than from experiments, thus failing condition (4).  Relying on simulated \C{$(x,z)$} pairs incurs the computational cost we seek to avoid, and limits the role of these fast simulators to supplementing traditional \E{simulators}, rather than replacing them. 

An alternative class of unsupervised, generative ML models are variational autoencoders (VAEs). While GANs leverage an adversarial training scheme, VAEs instead optimize a variational bound on the data's likelihood by constructing an intermediate latent space, $\mathcal{Y}$, which is distributed according to a prior, $p(y)$ \cite{zhang2018advances}.
An \E{encoder} ($\mathcal{X} \rightarrow \mathcal{Y}$) network transforms $x \rightarrow \tilde{y}$, where the \textasciitilde  ~distinguishes \C{a mapped sample} from those drawn from $p(y)$. Similarly, a \E{decoder} ($\mathcal{Y} \rightarrow \mathcal{X}$) network transforms \C{a sample} produced by the encoder back to the data space, $\tilde{y} \rightarrow \tilde{x}$. The autoencoder structure is the combined  encoder-decoder chain, $x\rightarrow \tilde{y} \rightarrow \tilde{x}$.
During training, the distribution of the encoder output,  $p_E(y \mid x)$, is constrained to match the latent space prior, $p(y)$, via a latent loss term which measures the distance between the distributions. At the same time, the output of the autoencoder, $\tilde{x}$, is constrained to match the input, $x$, which are compared pairwise. New samples from $\mathcal{X}$ following the distribution of the data, $p(x)$, can then be produced by decoding samples, \C{$\{y\}$}, drawn from $p(y)$, \C{via} $y \rightarrow \tilde{x}'$. 

The form of $p(y)$ is usually independent of the nature of the \C{problem's underlying theoretical model}, and is often chosen to be a multi-dimensional Gaussian for simplicity. This choice provides sufficient expressive power even for complex datasets (i.e. natural images). However, in the particle physics community, optimizing the encoding mapping to match this latent space is seen as an extra, unnecessary hurdle in training~\cite{ganlhc}. Therefore, GANs have been largely favored over VAEs in the pursuit of a fast particle physics simulator. Some studies investigated  VAEs in this context, but retained the unphysical form of $p(y)$ \C{(i.e. multi-dimensional Gaussian)}~\cite{DIBGANVAE, highVAE, e2eSAE}, preventing them from being conditional generators, failing requirement (2).

\section{\CC{Proposed Solution}} \label{sec:solution} 

\subsection{Our Approach: OTUS} \label{subsec:approach}
\CC{In this work, we aim to align the probabilistic autoencoder's latent space, $\mathcal{Y}$, with that of our inference task, $\mathcal{Z}$. This will allow us to learn a conditional simulation mapping from our theoretical model latent space to our data space, $\mathcal{Z} \rightarrow \mathcal{X}$.}
\CC{Therefore, we construct a probabilistic autoencoder where the latent space prior, $p(y)$, is identical to the physical latent space, $p(y) \equiv p(z) = p(z \mid \theta)$, for the choice of particular parameters, $\theta$.} 
The decoder then learns $p_D(x \mid z)$ providing precisely the desired conditional transformation, $z \rightarrow x$. Additionally, $p_D(x \mid z)$ can act as a tractable transfer function in approaches which estimate $p(x \mid \theta)$ via direct integration~\cite{Abazov_2004}. The encoder's learned $p_E(z \mid x)$ is of similar interest in unfolding applications~\cite{unfold_INN, unfold_OmniFold}. 

This is not possible with VAEs because optimizing the variational objective requires explicit computation of the densities $p(y), p_E(y \mid x)$, and $p_D(x \mid y)$. Therefore, $p(y)$ is often assumed to be a standard isotropic Gaussian for its simplicity and potential for uncovering independent latent factors of the data generation process. 
However, in particle physics the true prior, $p(z)$, which is governed by quantum field theory, is highly non-Gaussian and computing its density explicitly requires an expensive numerical procedure. 
Similarly, as we have little knowledge about the true underlying stochastic transforms, assuming any particular parametric density model for $p_E(y \mid x)$ or $p_D(x \mid y)$, like a multivariate Gaussian, would be inappropriate and overly restrictive. These concerns led us to use inherently stochastic (i.e. \E{implicit}) models for $p(z), p_E(z \mid x)$, and $p_D(x \mid z)$ that are fully \E{sample-driven}. 

Additionally, the VAE objective's use of KL-divergence introduces technical disadvantages. The KL-divergence, $\E{D_{\rm KL}}(\cdot  \| \cdot)$, is not a true distance metric, and will diverge for non-overlapping distributions often leading to unusable gradients during training~\cite{swae,WGAN}. Moreover, the specific use of $\E{D_{\rm KL}}(p_E(z \mid x)\| p(z))$ within the VAE loss forces $p_E(z \mid x)$ to match $p(z)$ for \E{ every} value of $x \sim p(x)$~\cite{wae}. This term must be carefully tuned (e.g. with a $\beta$-VAE approach \cite{betaVAE,DIBGANVAE}) to avoid the undesirable effect of the encoder mapping different parts of $\mathcal{X}$ to the same overlapping region in $\mathcal{Z}$, which \E{can be} particularly problematic if $\mathcal{Z}$ represents a physically meaningful latent space.

We resolve these issues by applying an emerging class of probabilistic autoencoders, based instead on the Wasserstein distance, which is a well-behaved distance metric between arbitrary probability distributions rooted in concepts from optimal transport theory~\cite{wae, swae}. 

The  original Wasserstein Autoencoder (WAE)~\cite{wae} loss function is

\begin{align} \label{eq:wae}
    \E{\mathcal{L}_{\rm WAE}}(p(x), p_D(x \mid z), p_E(z \mid x)) =&  ~ \underset{\ref{eq:wae}A}{\mathbb{E}_{x \sim p(x)} \mathbb{E}_{p_E(z \mid x)} \mathbb{E}_{\tilde{x} \sim p_D(x \mid z)} [c(x,  \tilde{x})]} 
    + \lambda ~ \underset{\ref{eq:wae}B}{d_z(p_E(z), p(z))}, 
\end{align}

\noindent where $\mathbb{E}$ denotes the expectation operator and $c(\cdot, \cdot)$ is a cost metric. For the optimal $p_E(z \mid x)$, \E{$\mathcal{L}_{\rm WAE}$} becomes an upper bound on the Wasserstein distance between the true data distribution, $p(x)$, and the decoder's learned distribution, $p_D(x)= \int dz p_D(x \mid z) p(z)$; the bound is tight for deterministic decoders.

Term A of \E{Equation} (\ref{eq:wae}) constrains the output of the encoder-decoder mapping, $\tilde{x}$, to match the input, $x$, while term B of \E{Equation} (\ref{eq:wae}) constrains the encoder mapping.
The hyperparameter $\lambda$ provides a relative weighting between the two terms. The difference between the marginal encoding distribution, $p_E(z)=\int dx p_E(z \mid x) p(x)$, and the latent prior, $p(z)$, is measured by $d_z(\cdot, \cdot)$.%
\footnote{Comparing $p_E(z)$ and $p(z)$ rather than $p_E(z \mid x)$ and $p(z)$ is the crucial innovation which allows different parts of $\mathcal{Z}$ to remain disjoint.} Unfortunately, the originally proposed options for $d_z(\cdot, \cdot)$~\cite{wae} had undesirable features which made them ill-suited for this particle physics problem \C{(see \nameref{subsec:modelchoice})}.

The more recent Sliced Wasserstein Autoencoder (SWAE)~\cite{swae} uses the Sliced Wasserstein (SW) distance as the $d_z(\cdot, \cdot)$ metric. The SW distance, $\E{d_{\rm SW}}(\cdot, \cdot)$, is a rigorous approximation to the Wasserstein distance, $d_{W}(\cdot, \cdot)$. The SWAE completely grounds the loss function in optimal transport theory as each term and the total loss can be identified as approximating the Wasserstein distances between various distributions and allows $p(y)$ to be \E{any} sampleable distribution, including the physical, $p(z)$. Additionally, the (S)WAE method allows the encoder and decoder to be implicit probability models, while avoiding an adversarial training strategy which can lead to problems like mode collapse~\cite{GANmodecollapse}. 

Both $d_W$ and $\E{d_{\rm SW}}$ are true distance metrics~\cite{swae}. The KL-divergence and adversarial schemes lack this property resulting in divergences and meaningless loss values which lead to problems during training and make it difficult to include additional, physically-motivated constraints. The Wasserstein distance is the cost to transport probability mass from one probability distribution to another according to a cost metric, $c(\cdot, \cdot)$, following the optimal transportation map. However, it is difficult to calculate for multivariate probability distributions when pairs from the optimal transportation map are unknown. 
However, for univariate probability distributions, there is a closed-form solution involving the difference between the inverse Cumulative Distribution Functions ($\E{{\rm CDF}}^{-1}$s) of the two probability distributions. The SW distance approximates the Wasserstein distance by averaging the one-dimensional Wasserstein distance over many randomly selected slices --- one-dimensional projections of the full probability distribution~\cite{swae} (see \nameref{subsec:training}).

The SWAE loss takes the general form of the WAE loss 

\begin{align} \label{eq:swae}
    \E{\mathcal{L}_{\rm SWAE}}(p(x), p_D(x \mid z), p_E(z \mid x)) =& ~ \underset{\ref{eq:swae}A}{\mathbb{E}_{x \sim p(x)} \mathbb{E}_{p_E(z \mid x)} \mathbb{E}_{\tilde{x} \sim p_D(x \mid z)} [c(x,  \tilde{x})]}
    + \lambda ~ \underset{\ref{eq:swae}B}{\E{d_{\rm SW}}(p_E(z), p(z))}.
\end{align}

\noindent Term A of \E{Equation} (\ref{eq:swae}) compares pairs \C{$(x,\tilde{x})$}, where $\tilde{x}$ is the output of the encoder-decoder mapping. In term B of \E{Equation} (\ref{eq:swae}), matched pairs are not available so we instead use the SW distance approximation. Both loss terms use the cost metric $c(u,v)=||u-v||^2$~\cite{swae}.

The SWAE allows us to train a probabilistic autoencoder that transforms between $\mathcal{X}$ and $\mathcal{Z}$ with a physical prior $p(z)$. However, since we are in an unsupervised setting, the true $p(x \mid z)$ is unknown. It is therefore crucial to ensure that the learned transformation is plausible and represents a series of physical interactions. To encourage this, we can easily impose supplemental physically-meaningful constraints on the SWAE model. These constraints can be relations between $\mathcal{Z}$ and $\mathcal{X}$ spaces or constraints on the internal properties of these respective spaces. In this work, we use one constraint from each category.

From the first category, we add a term comparing the unit vector parallel to the momentum of an easily identifiable particle in the latent and experimental spaces. This can be thought of as analogous to choosing a consistent basis and can be helpful for problems containing simple inversion symmetries\C{. An example of such an inversion symmetry exists in the $Z\rightarrow e^+e^-$ study below. In particle experiments, misidentification of lepton charge in the process of data reconstruction is known to be extremely rare. This means a learned mapping which frequently maps electron/positron ($e^\mp$) information in $\mathcal{Z}$ to positron/electron ($e^\pm$) information in $\mathcal{X}$, and vice versa, would be unphysical.}. For a generative mapping $G: \E{\mathcal{U}} \rightarrow \E{\mathcal{V}}$, this \E{anchor} term takes the general form

\begin{align}
    \mathcal{L}_{A} (p(u), p_G(v \mid u)) = ~ \mathbb{E}_{u \sim p(u)} \mathbb{E}_{v \sim p_G(v \mid u)} [c_A(u, v)] \label{eq:anchor}.
\end{align}

\noindent We chose $c_A(u, v) = 1 - \hat{\textbf{p}}_{u} \cdot \hat{\textbf{p}}_{v}$, where $\hat{\textbf{p}}$ is the unit vector of the electron's momentum. We add the anchor loss in $\mathcal{Z}$ space, $\mathcal{L}_{A} (p(x), p_E(z \mid x))$, and in $\mathcal{X}$ space,  $\mathcal{L}_{A} (p(z), p_D(x \mid z))$, to the SWAE loss with hyperparameter weightings $\beta_E$ and $\beta_D$ respectively.  

From the second category, we enforce the Minkowski metric constraint internally for $\mathcal{Z}$ and $\mathcal{X}$ spaces respectively. A particle's nature, excluding discrete properties such as charge and spin, is described by four quantities related by the Minkowski metric. Arranging these quantities into a 4-vector defined as $p^\mu = (\mathbf{p}, E)$ where $E$ is a particle's energy and $\mathbf{p}$ is a vector of its momentum in the \E{$\hat{\mathbf{x}}$, $\hat{\mathbf{y}}$, $\hat{\mathbf{z}}$} direction respectively, the constraint becomes

\begin{align}
    p^\mu p_\mu = E^2 - \mathbf{p}^2 = m^2,
\end{align}

\noindent where $m$ is the particle's mass. We directly enforce this relationship in the model for all particles.%
\footnote{We note that initial experiments lacked this constraint yet the networks automatically learned this relationship from the data. However, directly including this constraint in the model architecture improved performance overall.}

Adding more physically-motivated constraints would be straightforward, however, in this work we only assume this minimal set and recommend that more robust data structures be considered first, as such constraints may become unnecessary (see \nameref{sec:discussion}).

\C{
\subsection{OTUS in Practice}
In this section we briefly outline how OTUS might eventually be applied to problems in particle physics such as searches for new particles. However, we emphasize that this work only demonstrates a proof-of-principle version of OTUS. Follow-up work will be necessary to overcome some technical hurdles before OTUS could be applied to such a problem (see~\nameref{sec:discussion}).

A main goal of particle physics is to discover the complete set of fundamental units of matter: particles. Therefore, searches for exotic particles are common practice in this field. These searches typically proceed by looking for anomalies in data which are better described by simulations which assume the existence of a new particle. It is therefore phrased as a hypothesis test between two theoretical models, $\theta_{\rm SM}$, which assumes only the particles in the Standard Model (SM), and $\theta_{\rm BSM}$, which assumes the existence of one or more new particles that lie Beyond the Standard Model (BSM). These distinct models will generate distinct latent signatures, $\{z_{\rm SM} \mid \theta\}$ and $\{z_{\rm BSM} \mid \theta\}$, which lie in $\mathcal{Z}$. As particle physics experiments do not observe the latent $\{z\}$ directly, the hypothesis test is performed in the observed space $\mathcal{X}$, see \nameref{sec:intro} and \nameref{sec:objective} for more details.

\begin{figure}
    \centering
    \includegraphics[width=1.\linewidth]{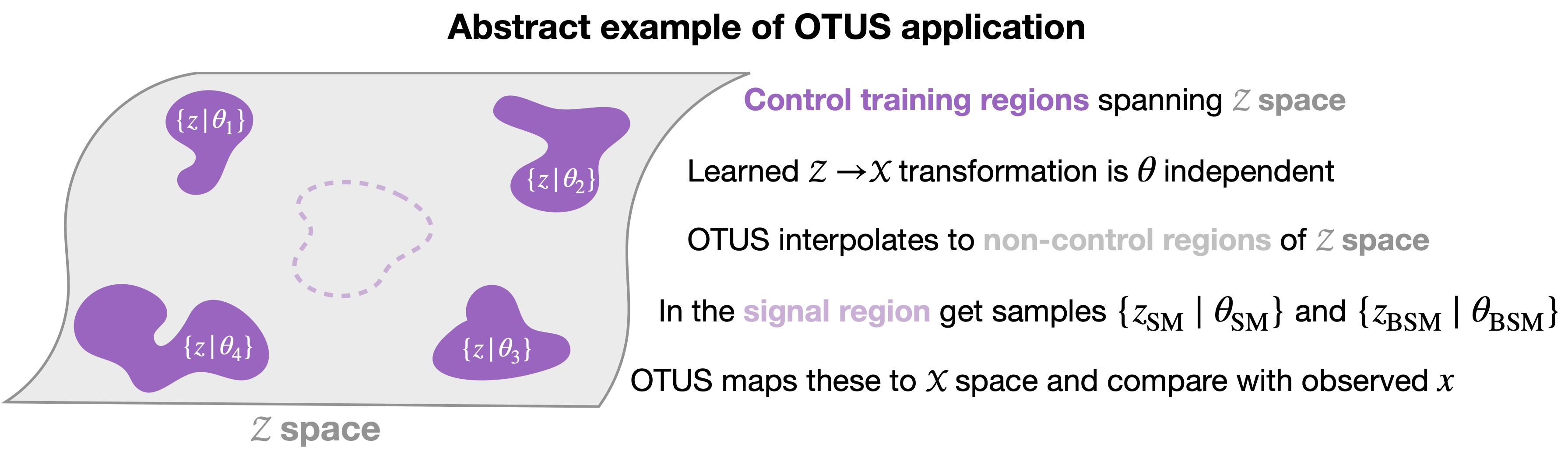}
    \caption{\C{ \textbf{Schematic diagram of how OTUS can be used in an abstract analysis.} 
    The gray surface represents $\mathcal{Z}$. Different theoretical models, $\theta_i$, will produce different signatures $\{z_i \mid \theta_i\}$ which lie in $\mathcal{Z}$. The goal of OTUS is to learn a general mapping from $\mathcal{Z} \rightarrow \mathcal{X}$ which is independent of the underlying theory, $\theta$, and only depends on the information contained in $\{z \in \mathcal{Z}\}$. One trains OTUS using control region data which span $\mathcal{Z}$ and have known outcomes in $\mathcal{X}$. These allow us to pair distributions in $\mathcal{Z}$ with distributions in  $\mathcal{X}$. From these examples, OTUS interpolates to the rest of $\mathcal{Z}$ and can then be used to generate $\{x_i\}$ from samples $\{z_i \mid \theta_i\}$ from regions not used during training, including the blinded signal region. This can then be used to search for new particles.}}  
    \label{fig:otus_example}
\end{figure}

The goal for OTUS is to learn a simulation mapping from $\mathcal{Z} \rightarrow \mathcal{X}$ which is independent of the underlying model, $\theta$, and can be applied to any $z$. This is achieved by carefully selecting control regions, $\{z_i \mid \theta_i\}$ , which span $\mathcal{Z}$ and for which observed data, $\{x\}$, is available for training. See Fig.~\ref{fig:otus_example} for a visual description. These control regions have known distributions of outcomes in $\mathcal{X}$, which allows us to properly match distributions in $\mathcal{Z}$ to distributions in $\mathcal{X}$ for training OTUS. Since these control regions are chosen to span $\mathcal{Z}$, OTUS will then be able to interpolate to unseen signal regions. Neural networks in general are known to perform well at interpolation tasks~\cite{NNinterpolation}, and recent work has shown that autoencoders in particular are proficient at learning manifold interpolation~\cite{Batson_2021}. Still more work has suggested there might be a deeper connection to the structure of this manifold and optimal transport~\cite{Komiske_2020}. Therefore, it is reasonable to expect that OTUS will be able to interpolate well in this space. However, these claims should be thoroughly investigated in future work.

A signal region is a region in $\mathcal{Z}$ space where signatures of new particles might occur. SM predictions, $\{z_{\rm SM} \mid \theta_{\rm SM}\}$, and BSM predictions, $\{z_{\rm BSM} \mid \theta_{\rm BSM}\}$, would then be passed to OTUS to produce two simulated data samples $\{x_{\rm SM}\}$ and $\{x_{\rm BSM}\}$ which would be compared with observed data, $\{x\}$, via a hypothesis test to calculate the relative likelihood of the SM and BSM theories. This technique, simulation-based inference, is standard practice in particle physics and is applied to existing simulation methods.

As a concrete example, let our BSM theory be the SM with the addition of a new particle, $Z'$, with a mass of 0.030 [TeV${\rm c}^{-2}$], which decays into a pair of leptons, a flagship search for the Large Hadron Collider~\cite{Aaij_2020}.  The latent space $\mathcal{Z}$ would include the two leptons produced by the decay of the $Z'$, and the observed space $\mathcal{X}$ would include the leptons identified and measured by the detector. For OTUS to be able to predict the observed signatures from this latent space, it would need to interpolate between control regions which have similar relationships.  Decays of existing particles to leptons, such as the 0.091 [TeV${\rm c}^{-2}$] $Z$ and the 0.002 [TeV${\rm c}^{-2}$] $J/\psi$ would allow OTUS to learn the mapping from latent leptons to observed leptons. Our theoretical $Z'$ has a mass which lies between those of the particles in our control regions. OTUS would need to interpolate along this axis; control regions at various masses provided by the $Z$ and $J/\psi$ decays are therefore essential to describe and determine the nature of the interpolation. To verify the interpolation, one might compare the prediction of OTUS to observed data in the intermediate range between the $Z'$ and the $Z$.

Alternatively, the $Z'$ could have a heavier mass, e.g. 1 [TeV${\rm c}^{-2}$]. In this scenario, OTUS would be required to extrapolate along the mass axis. Naively, this sounds problematic as extrapolation is generally much less sound than interpolation, however this task is also required of current simulations for this scenario. Simulations succeed in such tasks when they have inductive biases which control their behavior even outside of training (tuning) regions. These inductive biases are based on physics principles and scale to the signal regions of interest. For neural networks, it has been shown that architectures with inductive bias constraints succeed at such extrapolation tasks~\cite{GNsym}. Since a mature version of OTUS will manifestly include such inductive biases (see \nameref{sec:discussion}) it is reasonable to assume it can achieve this task as well as current simulation methods can.
}

\section{Results} \label{sec:results_} 

\subsection{Demonstration in \texorpdfstring{$Z \rightarrow e^+e^-$}{Z -> e+ e-} decays} \label{subsec:demoZee}
We first test OTUS on an important control region: leptonic decays of the $Z$-boson to electron-positron pairs, $Z\rightarrow e^+e^-$. The theoretical prior is well-known, and its parameters \C{$\{\theta\}$}, like the  $Z$-boson's mass and its interaction strengths, are tightly constrained by precision experiments.  We identify $\mathcal{Z}$ with the $Z$-boson's decay products: the electron, $e^-$, and positron, $e^+$, whose four-momenta span the space.  We compose these into an eight-dimensional vector

\begin{align}
    z := \{ z_{e^-},z_{e^+} \} 
       = \{\textbf{p}^{e^-}, E^{e^-}, 
            \textbf{p}^{e^+}, E^{e^+} \}.
\end{align}

\noindent This simplistic vector description excludes categorical properties such as charge.

The model prior $p(z)$ can be simply expressed with quantum field theory and \C{sampled}.  The subsequent step, where the electron and positron travel through the layers of detectors, depositing energy and causing particle showers, cannot be described analytically; a model will be learned by OTUS from data in control regions. Here we use simulated data samples, but specific \C{$(z, x)$} pairs are not used to mimic the information available when training from real data. The complex intermediate state with many low-energy particles and high-dimensional detector readouts is reduced and reconstructed yielding estimates of the electron and positron four-momenta.  Therefore, $\mathcal{X}$ has the same structure and dimensionality as $\mathcal{Z}$, though the distribution $p(x)$ reflects the impact of the finite resolution of detector systems (see \nameref{subsec:datagen}).

\begin{figure}
    \centering
    \vspace{1cm}
    \begin{subfigure}{1.\textwidth}
    \includegraphics[width=\linewidth]{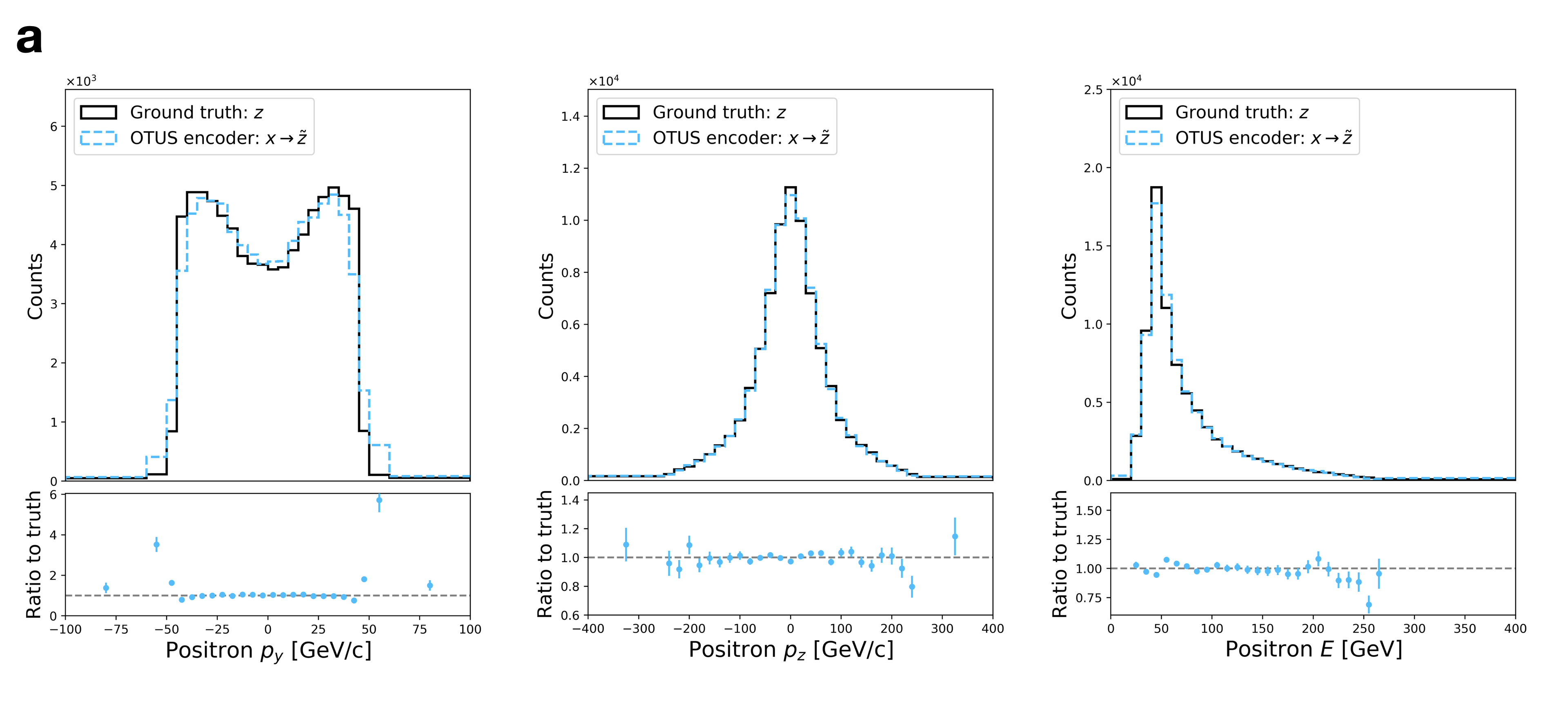}
    \end{subfigure}
    \begin{subfigure}{1.\textwidth}
    \includegraphics[width=\linewidth]{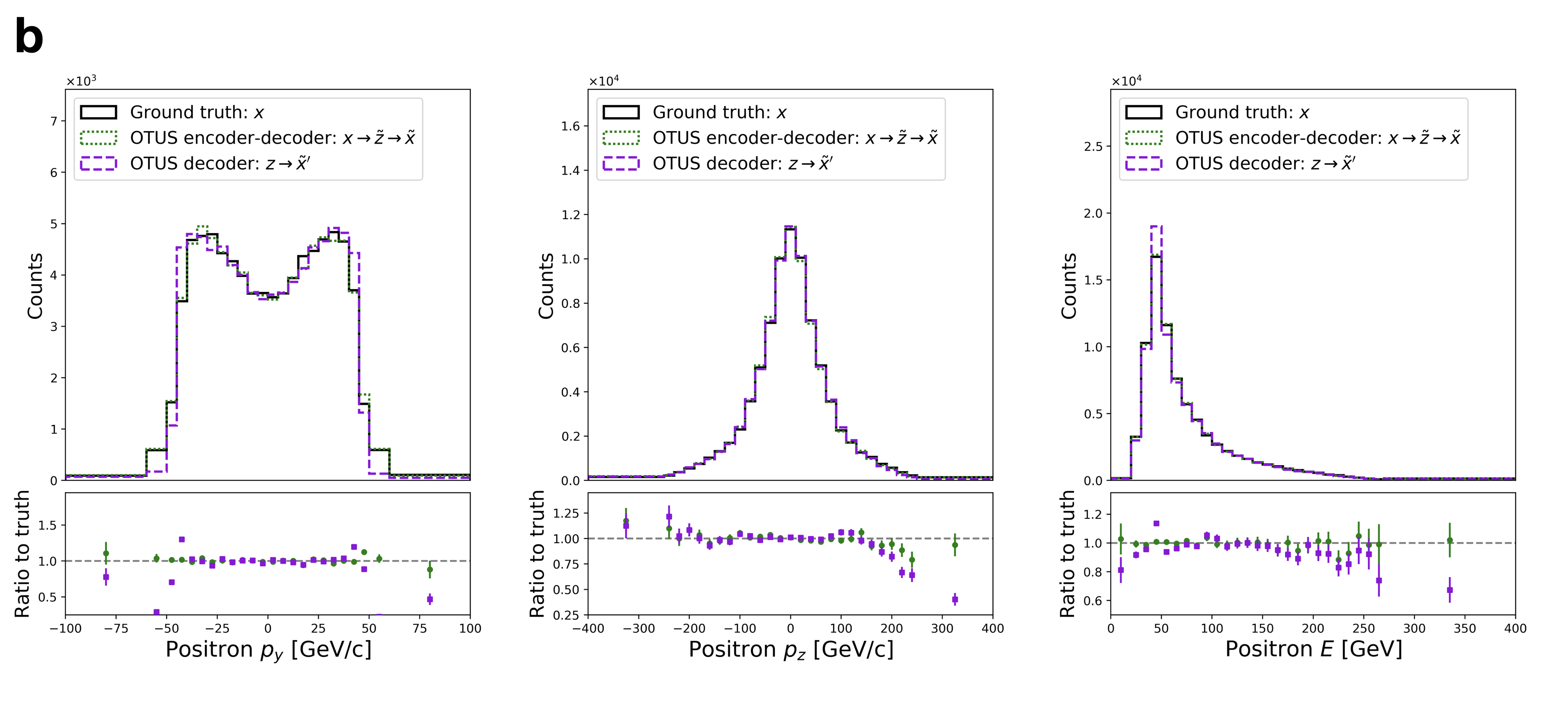}
    \end{subfigure}
    \caption{\E{\textbf{Performance of OTUS for $Z\rightarrow e^+ e^-$ decays}}. 
    \E{\textbf{a} Matching of the positron's $p_x$, $p_y$, and $E$ distributions in $\mathcal{Z}$.} \C{It} shows distributions of samples from the theoretical prior, \C{$\{z \sim p(z)\}$ (solid black)}, as well as the output of the encoder\C{, $\{\tilde{z}\}$; the encoder } transforms samples of testing data in experimental space, $\mathcal{X}$, to the latent space, $\mathcal{Z}$, \C{and is } shown as $x\rightarrow \tilde{z}$ \C{(dashed cyan)}.  
    \E{\textbf{b} Matching of the positron's $p_x$, $p_y$, and $E$ distributions in $\mathcal{X}$.} \C{It} shows the testing sample \C{$\{x \sim p(x)\}$ (solid black)} in the experimental space, $\mathcal{X}$, as well as output from the decoder applied to samples drawn from $p(z)$, labeled as $z\rightarrow \tilde{x}'$ \C{(dashed purple)}. Also shown are samples passed through both the decoder and encoder chain, $x\rightarrow\tilde{z}\rightarrow\tilde{x}$ \C{(dotted green)}. \C{Dotted green and solid black distributions are matched explicitly during training. Enhanced differences between dashed purple and solid black indicate the encoder's output needs improvement, as $p_E(z)$ does not fully match $p(z)$.}
    \C{If performance were ideal, the distributions in every plot would match up to statistical fluctuations.} 
    Residual plots show bin-by-bin ratios with statistical uncertainties propagated accordingly (see \nameref{subsec:eval}).}
    \label{fig:z_space}
\end{figure}

Figure~\ref{fig:z_space} shows distributions of testing data, unpaired samples from $\mathcal{X}$ and $\mathcal{Z}$ in several projections, and the results of applying the trained encoder and decoder to transform between the two spaces.  Visual evaluation indicates qualitatively good performance, and quantitative metrics are provided. Measuring overall performance, the SW distances are as follows: $\E{d_{\rm SW}}(p(z), p_E(\tilde{z})) = 0.984$ [$\operatorname{GeV}^2$],
$\E{d_{\rm SW}}(p(x), p_D(\tilde{x})) = 1.33$ [$\operatorname{GeV}^2$], 
$\E{d_{\rm SW}}(p(x), p_D(\tilde{x}')) = 3.03$ [$\operatorname{GeV}^2$]. 
Additionally, several common metrics are reported for each projection in Supplementary Tables 1 and 2. Details of the calculations are provided in \nameref{subsec:eval}.  

\begin{figure}
    \centering
    \begin{subfigure}{.45\textwidth}
   \includegraphics[height=7.55cm]{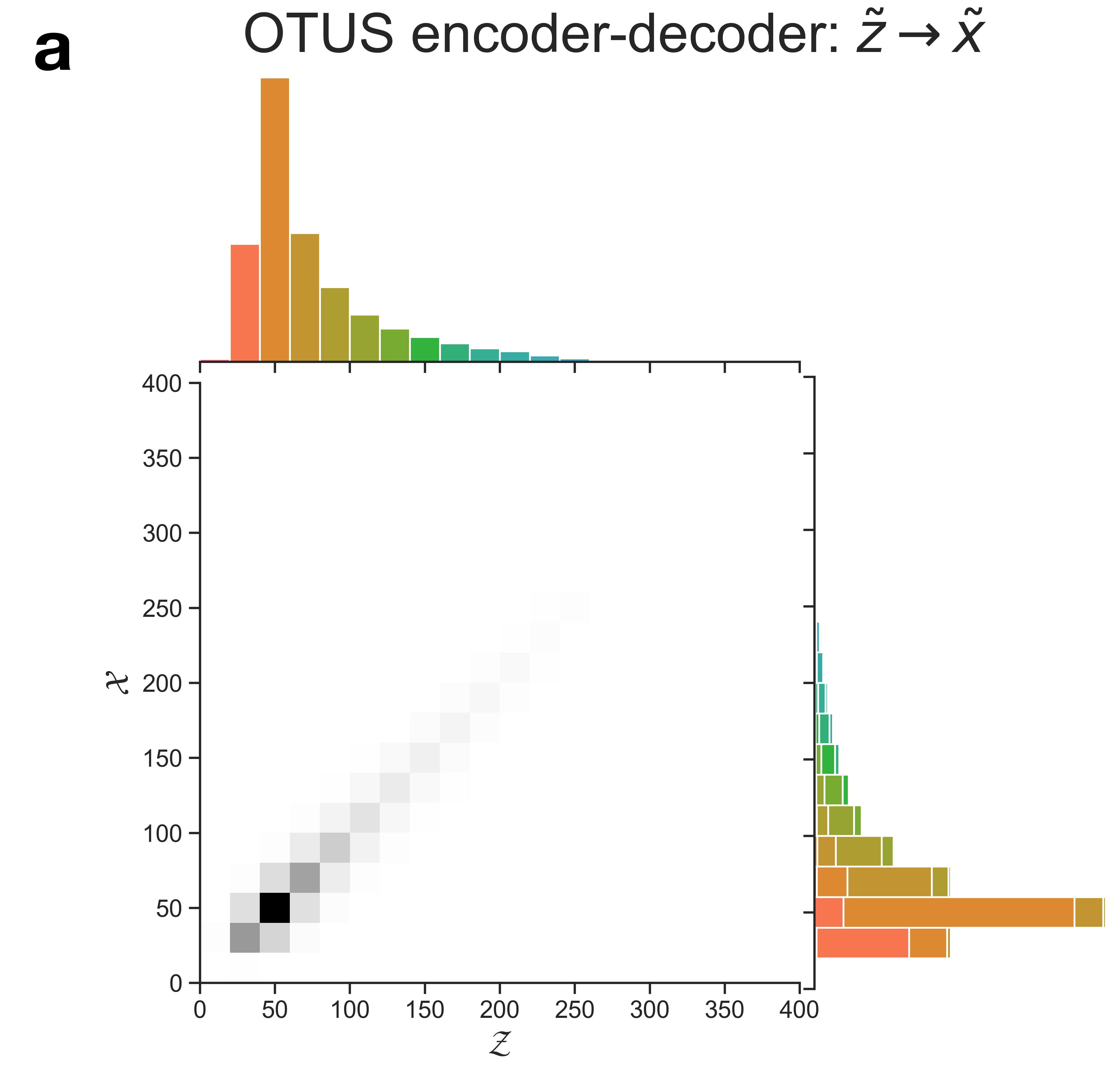}
    \end{subfigure}
    \begin{subfigure}{.45\textwidth}
    \includegraphics[height=7.55cm]{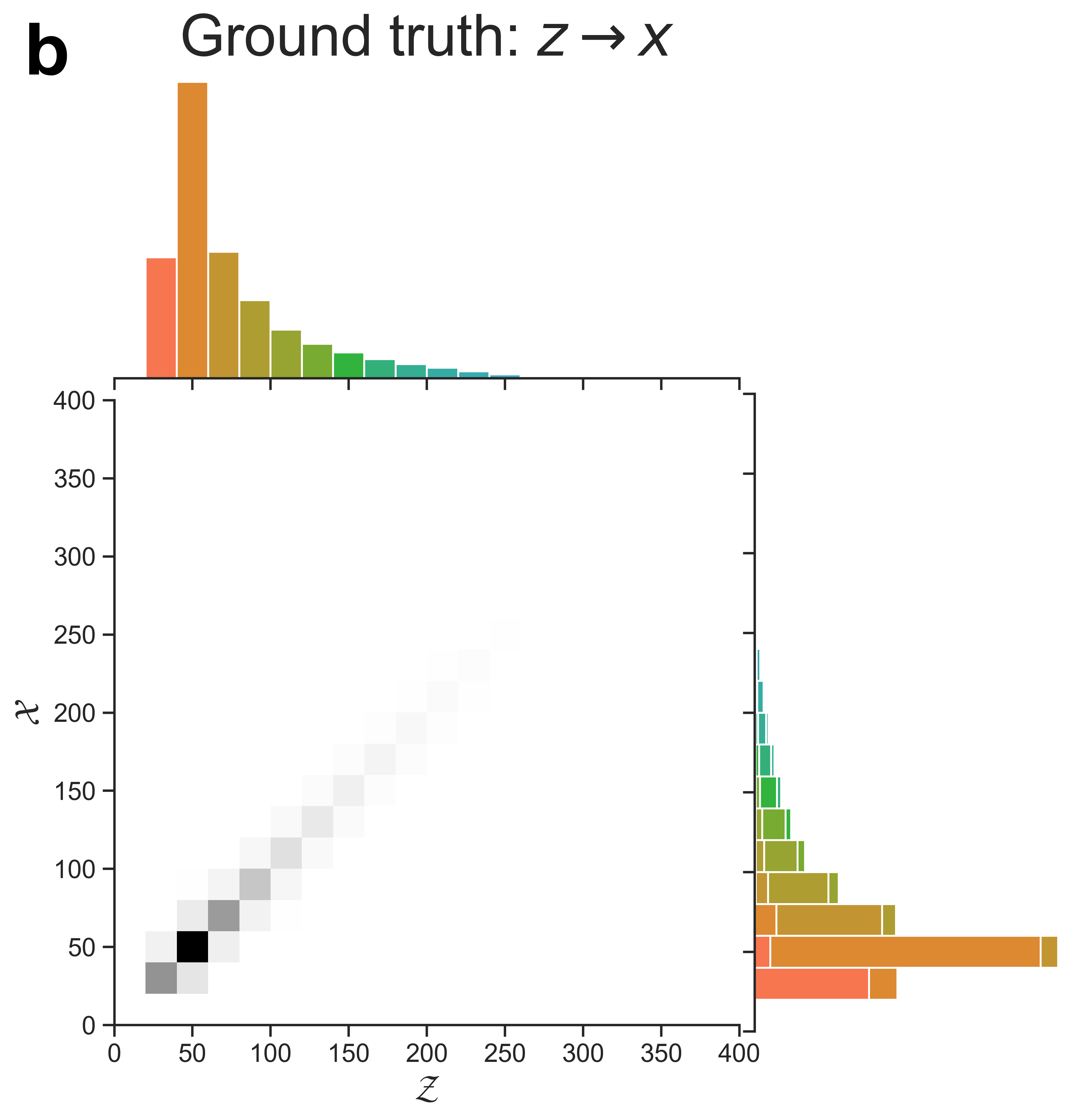}
    \end{subfigure}
    \caption{ \E{ \textbf{Visualization of the  transformation from $\mathcal{Z} \rightarrow \mathcal{X}$ in the $Z\rightarrow e^+ e^-$ study for positron energy.}}  
    \E{\textbf{a} The learned} transformation of the decoder, $p_D(x \mid z)$.  
    \E{\textbf{b} The true} transformation from the simulated sample, for comparison, though the true \C{$(z,x)$} pairs are not typically available and were not used in training.
    Colors in the $\mathcal{X}$ projection indicate the source bin in $\mathcal{Z}$ for a given sample.}
    \label{fig:z_trans}
\end{figure}

To ensure that the learned decoder reflects the physical processes being modeled, we inspect the transformation from $\mathcal{Z} \rightarrow \mathcal{X}$ in Figure~\ref{fig:z_trans}. The learned transfer function, $p_D(x \mid z)$, shows reasonable behavior, mapping samples from $\mathcal{Z}$ to nearby values of $\mathcal{X}$\C{. This reflects the imperfect resolution of the detector while avoiding unphysical transformations such as mapping information on the far-end distribution tails in $\mathcal{Z}$ to the distribution peaks in $\mathcal{X}$.}

\begin{figure}
    \centering
    \begin{subfigure}{0.45\textwidth}
    \includegraphics[width=.85\linewidth]{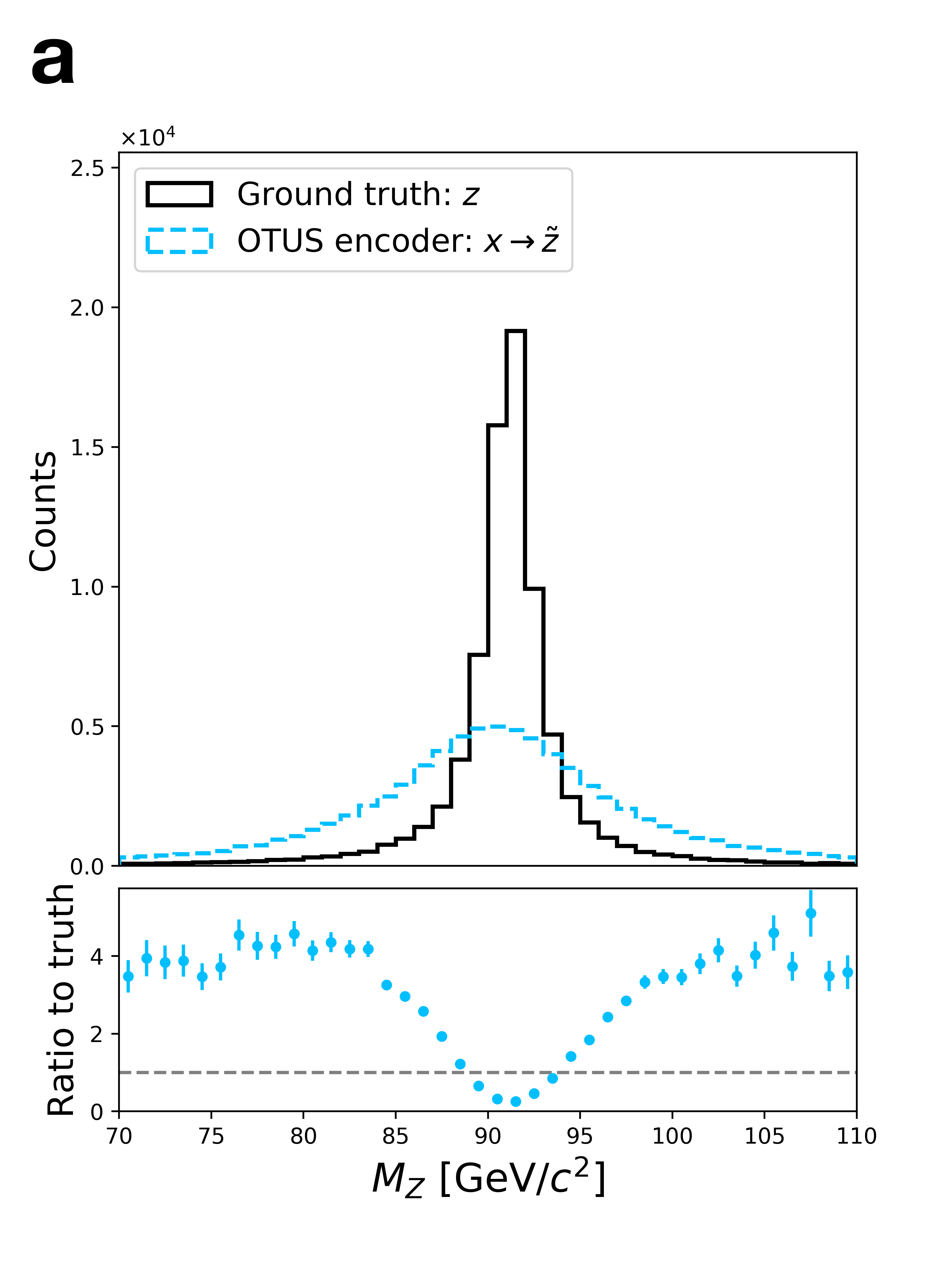}
    \end{subfigure}
    \begin{subfigure}{0.45\textwidth}
    \includegraphics[width=.85\linewidth]{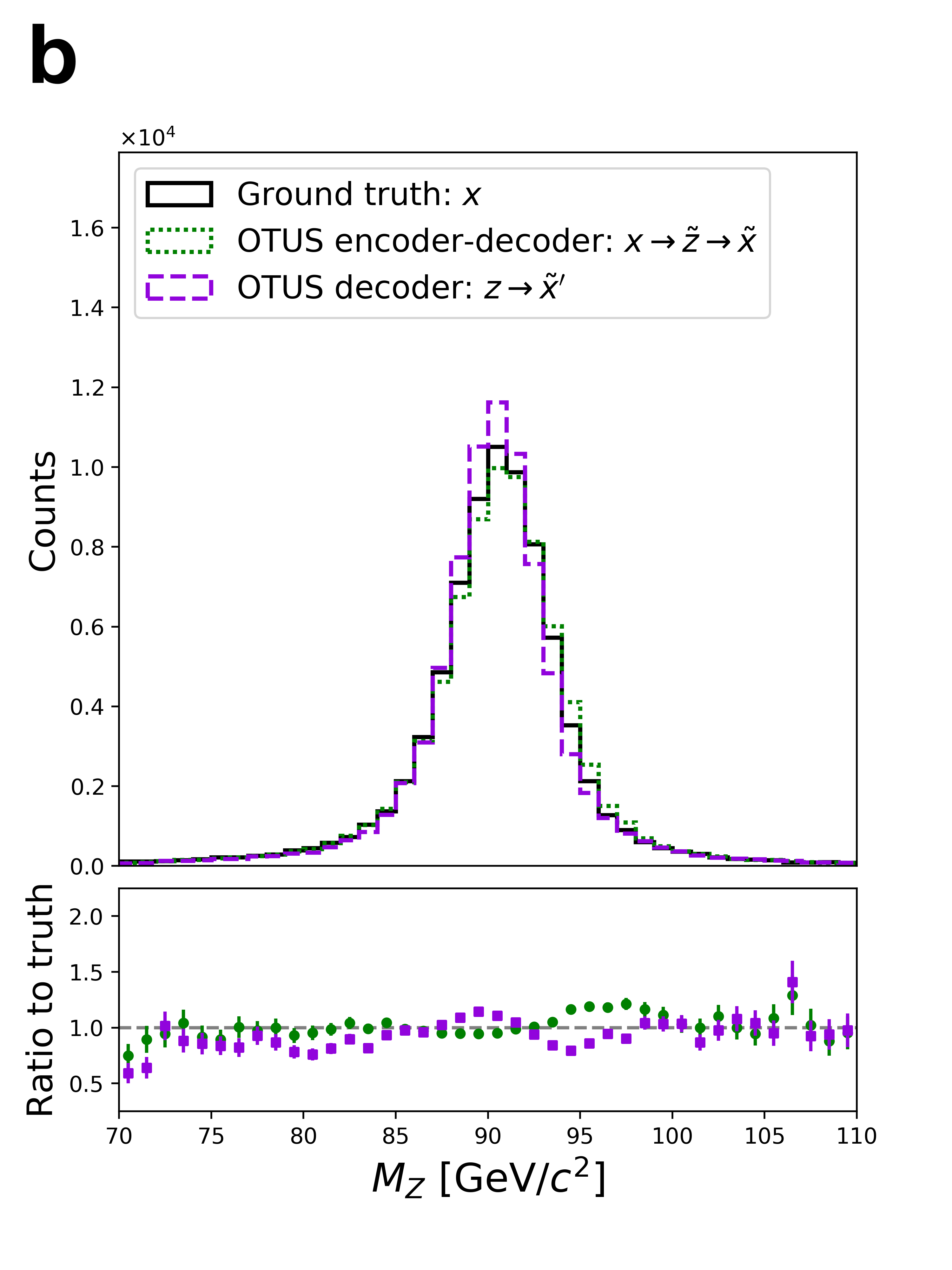}
    \end{subfigure}
    \caption{\E{ \textbf{Performance of OTUS for $Z\rightarrow e^+ e^-$ decays in a physically important derived quantity, the invariant mass of the electron-positron pair, $M_Z$.}} 
    \E{\textbf{a} Matching of the $M_Z$ distribution in $\mathcal{Z}$. It shows distributions of samples from the theoretical prior, $\{z \sim p(z)\}$ (solid black), as well as the output of the encoder, $\{\tilde{z}\}$; the encoder transforms samples of testing data in experimental space, $\mathcal{X}$, to the latent space, $\mathcal{Z}$, and is shown as $x\rightarrow \tilde{z}$ (dashed cyan).}
    \E{\textbf{b} Matching of the $M_Z$ distribution in $\mathcal{X}$. It shows the testing sample $\{x \sim p(x)\}$ (solid black) in the experimental space, $\mathcal{X}$, as well as output from the decoder applied to samples drawn from $p(z)$, labeled as $z\rightarrow \tilde{x}'$ (dashed purple). Also shown are samples passed through both the decoder and encoder chain, $x\rightarrow\tilde{z}\rightarrow\tilde{x}$ (dotted green). Dotted green and solid black distributions are matched explicitly during training. Enhanced differences between dashed purple and solid black indicate the encoder's output needs improvement, as $p_E(z)$ does not fully match $p(z)$.} 
    \C{If performance were ideal, the distributions in every plot would match up to statistical fluctuations.}
    Note that this projection was not explicitly used during training, but was inferred by the networks. Residual plots show bin-by-bin ratios with statistical uncertainties propagated accordingly (see \nameref{subsec:eval}).}
    \label{fig:z_mass}
\end{figure}

Finally, we examine the distribution of a physically important derived quantity, the invariant mass of the $Z$-boson, see Figure~\ref{fig:z_mass}. This quantity was not used as an element of the loss function, and so provides an alternative measure of performance.  The results indicate a high-quality description of the transformation from $\mathcal{Z}$ to $\mathcal{X}$. The performance of the transformation from $\mathcal{X}$ to $\mathcal{Z}$ is less well-described, likely because this relation is more strict in $\mathcal{Z}$ causing a sharper peak in the distribution. Such strict rules are difficult for networks to learn when not penalized directly or hard-coded as inductive biases, again signaling that a robust data representation will be crucial to improving performance \E{(see \nameref{sec:discussion})}.

\subsection{Demonstration in semileptonic \E{top-quark} decays} \label{subsec:demottbar}

The $Z$-boson control region is valuable for calibrating simulations of leptons such as electrons or muons, which tend to be stable and well-measured.  We next test OTUS on the challenging task of modeling the decay and detection of \E{top-quark} pairs featuring more complex detector signatures. This control region has more observed particles and introduces additional complexities: unstable particles decaying in flight, significantly degraded resolution relative to leptons, undetected particles, and a stochastically variable number of observed particles.

The initial creation of \E{top-quark} pairs, their \C{leading-order} decay $t ~\bar{t} \rightarrow W^+b ~ W^-\bar{b}$, and the subsequent \E{$W$-boson} decays are well-described using quantum field theory, so $p(z \mid \theta)$ can be \C{sampled}. We select the modes $W^- \rightarrow e^- ~\bar{\nu}_e$ and $W^+ \rightarrow u ~\bar{d}$ as examples and assign our latent space to describe the four-momenta of these six of particles:

\begin{align}
    z:= \{ z_{e^-}, ~z_{\bar{\nu}_e}, ~z_b, ~z_{\bar{b}}, ~z_u, ~z_{\bar{d}} \} 
     = \{\textbf{p}^{e^-},        E^{e^-},
        \textbf{p}^{\bar{\nu}_e}, E^{\bar{\nu}_e},
        \textbf{p}^{b},           E^{b},
        \textbf{p}^{\bar{b}},     E^{\bar{b}},
        \textbf{p}^{u},           E^{u},
        \textbf{p}^{\bar{d}},     E^{\bar{d}}
        \}
\end{align}

\noindent with a total of twenty-four dimensions. 

Unlike in the $Z\rightarrow e^+e^-$ study, the $\mathcal{X}$ space's structure is considerably different from that of the $\mathcal{Z}$ space.  While the electron $e^-$ is stable and readily identifiable, the other particles are more challenging.  The neutrino, $\bar{\nu}_e$, is stable, yet invisible to our detectors, providing no estimate of its direction or momentum; instead its presence is inferred using momentum conservation $\textbf{p}^{\nu} = -\sum \textbf{p}^{\textrm{observed}}$. Unfortunately, soft initial state radiation and detector inefficiencies also contribute to missing momentum. The aggregate quantity is labeled $\textbf{p}^{\textrm{miss}}$. The four quarks $\bar{b}$, $u$, $\bar{d}$ and $b$ are strongly-interacting particles each producing complex showers of particles that are clustered together into \E{ jets} to estimate the original quark momenta and directions.  Unfortunately, despite significant recent \C{progress~\cite{Fenton:2020woz, Erdmann_2019, Erdmann_2014}}, we cannot assume a perfect identification of the source particle in $\mathcal{Z}$ for a given jet observed in $\mathcal{X}$, causing significant ambiguity. 

Additionally, a complete description of the $\mathcal{Z}\rightarrow \mathcal{X}$ transformation should include the possibilities for the number of jets in $\mathcal{X}$ to exceed the number of quarks, due to radiation and splitting, or to fail to match the number of quarks, due to jet overlap or detector inefficiency. We leave this complexity for future work and restrict our $\mathcal{X}$ space to contain exactly four jets. 

The final complexity introduced in this study is the presence of a sharp lower threshold in transverse momentum, $p_\textrm{T}$. Experimental limitations require that jets with $p_{\textrm{T}} < 20$ \E{[GeV${\rm c^{-1}}$]} be discarded and therefore are not represented in the training dataset, as they would be unavailable in control region data. Mimicking this experimental effect, we directly impose this threshold on the decoder's output instead of the network learning it. Paralleling reality, such events are discarded before computing losses. This strategy requires modifications to both the model and training strategy (see \nameref{sec:methods}).

Our \C{experimental} data is the vector

\begin{align}
    x :=& \{ x_{e^-}, x_{\textrm{miss}}, ~x_{\textrm{jet}1}, ~x_{\textrm{jet}2}, ~x_{\textrm{jet}3}, x_{\textrm{jet}4} \} \\
    =& \{\textbf{p}^{e^-},  E^{e^-},
        \textbf{p}^{\textrm{miss}},  E^{\textrm{miss}},
        \textbf{p}^{\textrm{jet}1}, E^{\textrm{jet}1},
        \textbf{p}^{\textrm{jet}2}, E^{\textrm{jet}2},
        \textbf{p}^{\textrm{jet}3}, E^{\textrm{jet}3},
        \textbf{p}^{\textrm{jet}4}, E^{\textrm{jet}4}
        \},
\end{align}

\noindent with a total of twenty-four dimensions. If quark-jet assignment were possible, it would be natural to align the order of the observed jets with the order of their originating quarks in $\mathcal{Z}$ space. Lacking this information, it is typical to order jets by descending $|\textbf{p}_{\textrm{T}}| = \sqrt{p_x^2 + p_y^2}$, where jet 1 has the largest $|\textbf{p}_{\textrm{T}}|$. 

\begin{figure}
    \centering
    \vspace{1cm}
    \begin{subfigure}{1.\textwidth}
    \includegraphics[width=\linewidth]{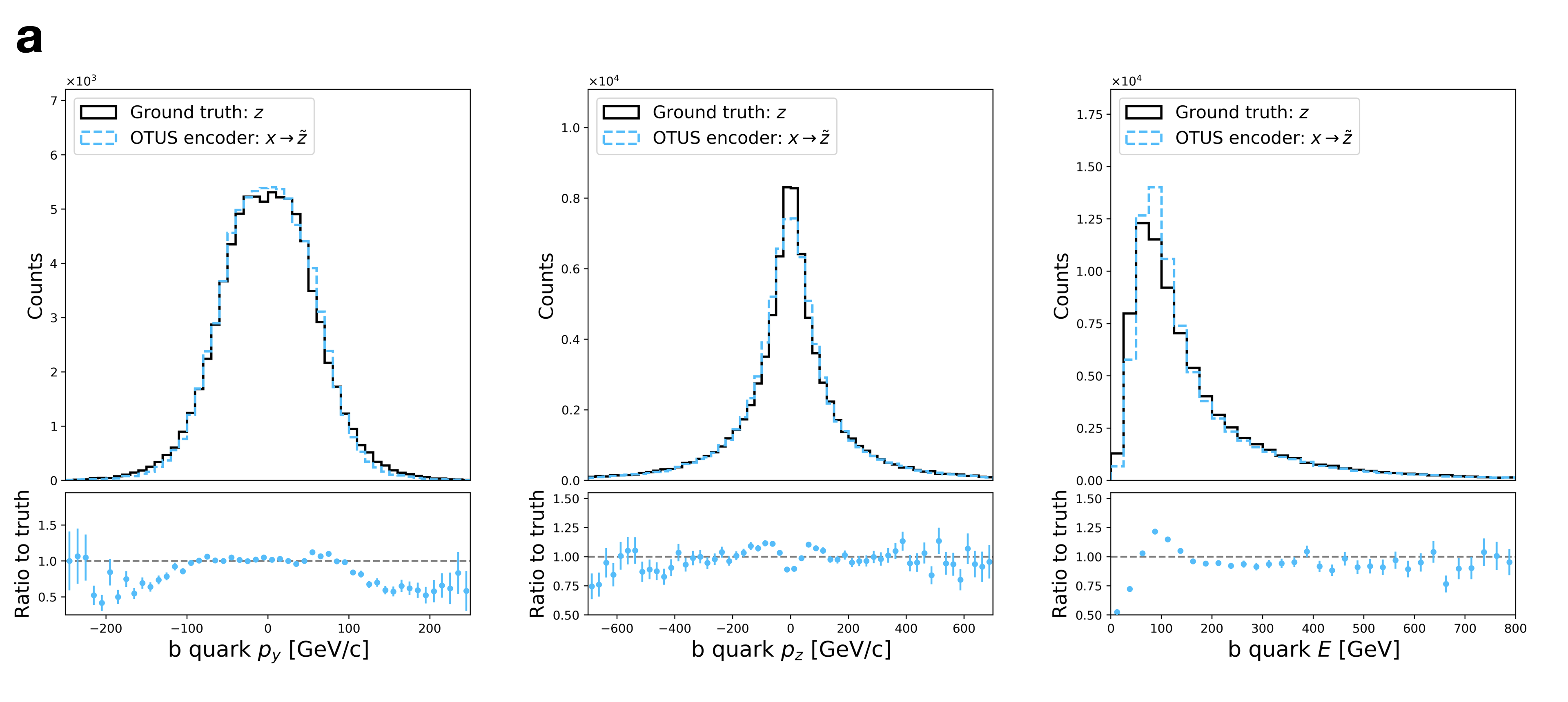}
    \end{subfigure}
    \begin{subfigure}{1.\textwidth}
    \includegraphics[width=\linewidth]{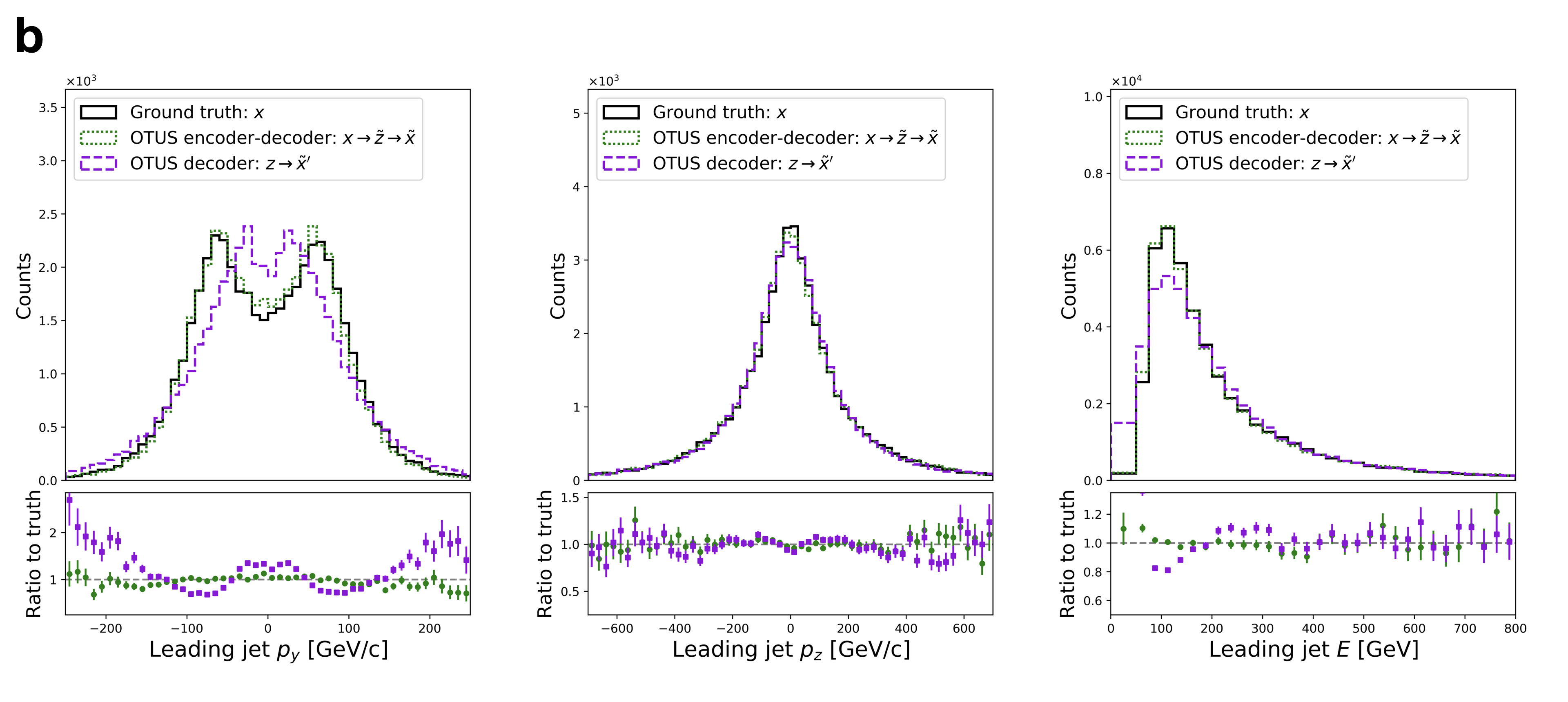}
    \end{subfigure}
    \caption{\E{\textbf{Performance of OTUS for semileptonic $t\bar{t}$ decays.}} 
    \E{\textbf{a} Matching of the $b$ quark's $p_x$, $p_y$, and $E$ distributions in $\mathcal{Z}$.} \C{It}
    shows distributions of samples from the theoretical prior, \C{$\{z \sim p(z)\}$ (solid black)}, as well as the output of the encoder\C{, \{$\tilde{z}\}$; the encoder} transforms samples of the testing data in experimental space, $\mathcal{X}$, to the latent space, $\mathcal{Z}$, \C{and is } shown as $x\rightarrow \tilde{z}$ \C{(dashed cyan)}.  
    \E{\textbf{b} Matching of the leading jet's $p_x$, $p_y$, and $E$ distributions in $\mathcal{X}$.} \C{It} shows the testing sample \C{$\{x \sim p(x)\}$ (solid black)} in the experimental space, $\mathcal{X}$, as well as output from the decoder applied to samples drawn from the prior $p(z)$, labeled as $z\rightarrow \tilde{x}'$ \C{(dashed purple)}. Also shown are samples passed through both the decoder and encoder chain, $x\rightarrow\tilde{z}\rightarrow\tilde{x}$ \C{(dotted green)}. \C{Dotted green and solid black distributions are matched explicitly during training. Enhanced differences between dashed purple and solid black indicate the encoder's output needs improvement, as $p_E(z)$ does not fully match $p(z)$.} 
    \C{If performance were ideal, the distributions in every plot would match up to statistical fluctuations.} 
    Residual plots show bin-by-bin ratios with statistical uncertainties propagated accordingly (see \nameref{subsec:eval}).
    }
    \label{fig:t_space}
\end{figure}

Figure~\ref{fig:t_space} shows distributions of testing data, unpaired samples from $\mathcal{X}$ and $\mathcal{Z}$ in several projections, and the results of applying the trained encoder and decoder to transform between the two spaces.   Visual evaluation indicates qualitatively good performance, and quantitative metrics are also provided. Measuring overall performance the SW distances are as follows: $\E{d_{\rm SW}}(p(z), p_E(\tilde{z})) = 22.3$ [$\operatorname{GeV}^2$],
$\E{d_{\rm SW}}(p(x), p_D(\tilde{x})) = 232$ [$\operatorname{GeV}^2$], 
$\E{d_{\rm SW}}(p(x), p_D(\tilde{x}')) = 120$ [$\operatorname{GeV}^2$]. 
Additionally, several common metrics are reported for each projection in Supplementary Table 3 and 4. Details of the calculations are provided in \nameref{subsec:eval}.   

\begin{figure}
    \centering
    \begin{subfigure}{.45\textwidth}
   \includegraphics[height=7.55cm]{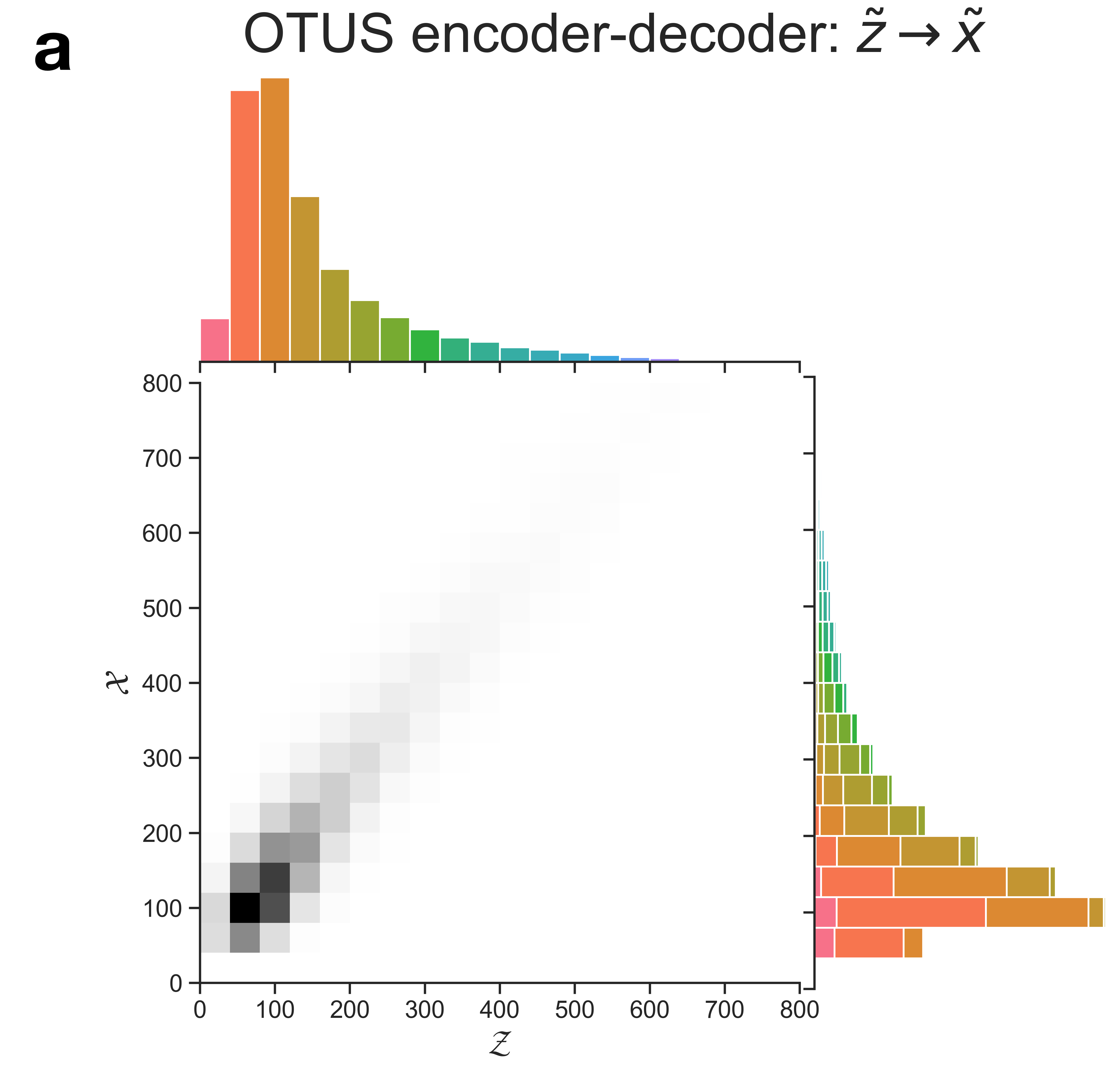}
    \end{subfigure}
    \begin{subfigure}{.45\textwidth}
    \includegraphics[height=7.55cm]{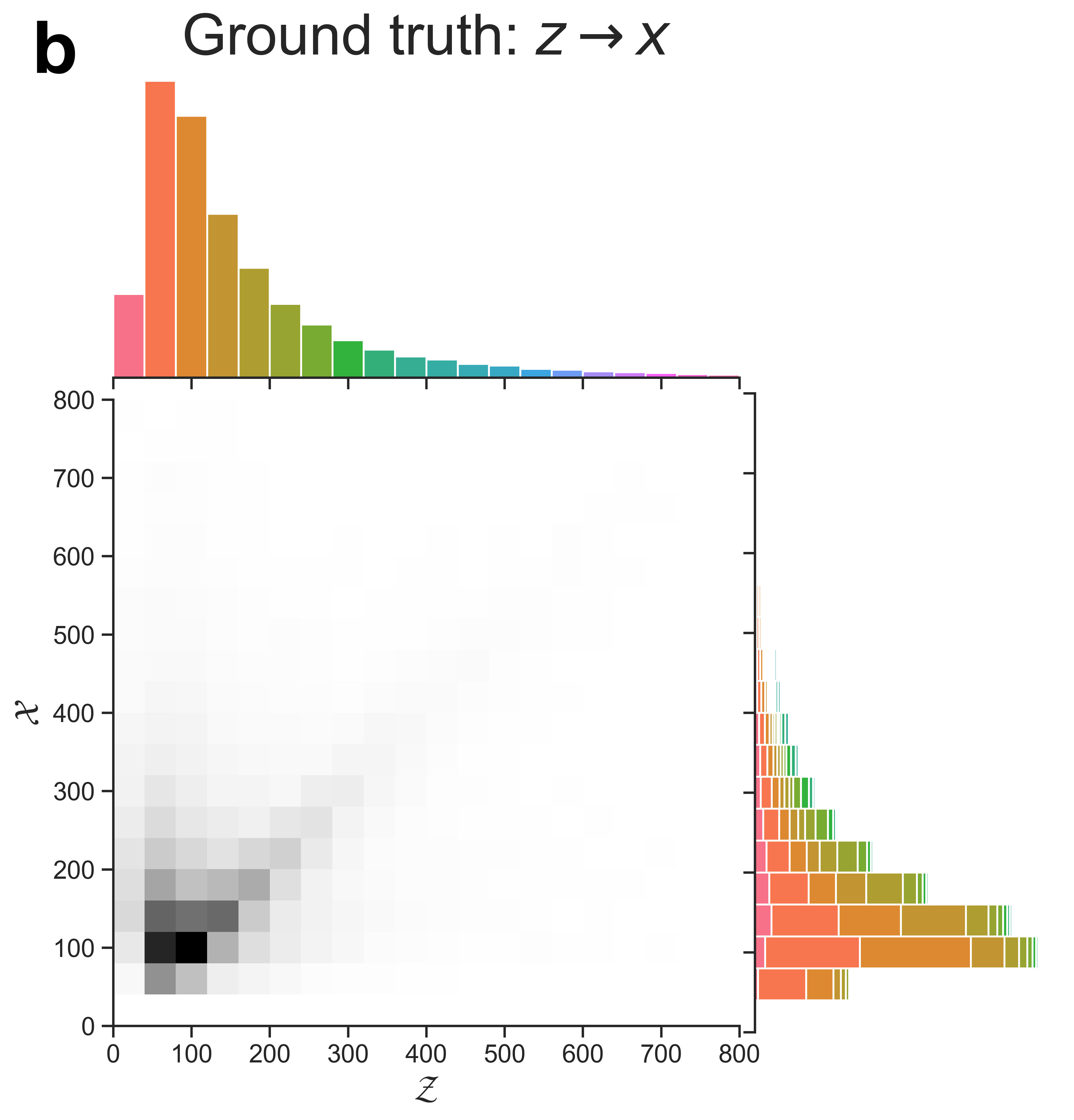}
    \end{subfigure}
    \caption{\E{ \textbf{Visualization of the  transformation from $\mathcal{Z} \rightarrow \mathcal{X}$ in the $t\bar{t}$ study for the energy of the $b$ quark in $\mathcal{Z}$ to energy of the leading jet in $\mathcal{X}$.}} 
    \E{\textbf{a} The learned} transformation of the decoder, $p_D(x \mid z)$.  
    \E{\textbf{b} The true} transformation from the simulated sample, for comparison, though the true \C{$(z,x)$} pairs are not typically available and were not used in training.
    Note that the $b$ quark will not always correspond to the leading jet, see the text for details. Colors in the $\mathcal{X}$ projection indicate the source bin in $\mathcal{Z}$ for a given sample.  }
    \label{fig:t_trans}
\end{figure}

To probe the $\mathcal{Z} \rightarrow \mathcal{X}$ transformation, we inspect the learned transfer function, $p_D(x \mid z)$ in Figure~\ref{fig:t_trans}. While the overall performance is worse in this more complex case, it still shows reasonable behavior, mapping samples from $\mathcal{Z}$ to nearby values of $\mathcal{X}$ and avoiding unphysical transformations\C{ such as mapping information on the far-end distribution tails in $\mathcal{Z}$ to the distribution peaks in $\mathcal{X}$.} Additionally, cross-referencing with the true simulation's mapping shows the similar nature of the mappings.

\begin{figure}
    \centering
    \begin{subfigure}{0.40\textwidth}
    \includegraphics[width=\linewidth]{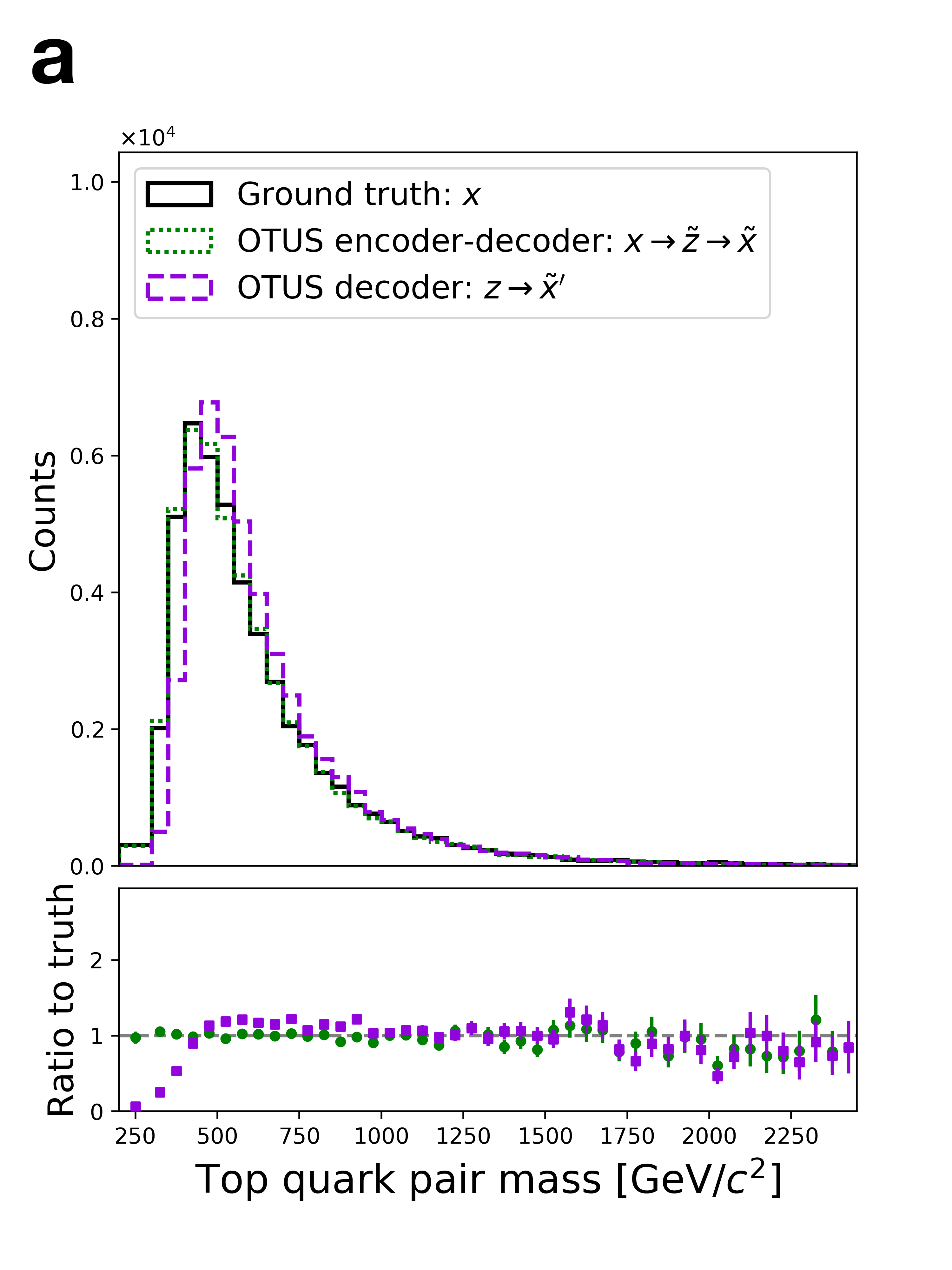}
    \end{subfigure}
    \begin{subfigure}{0.40\textwidth}
    \includegraphics[width=\linewidth]{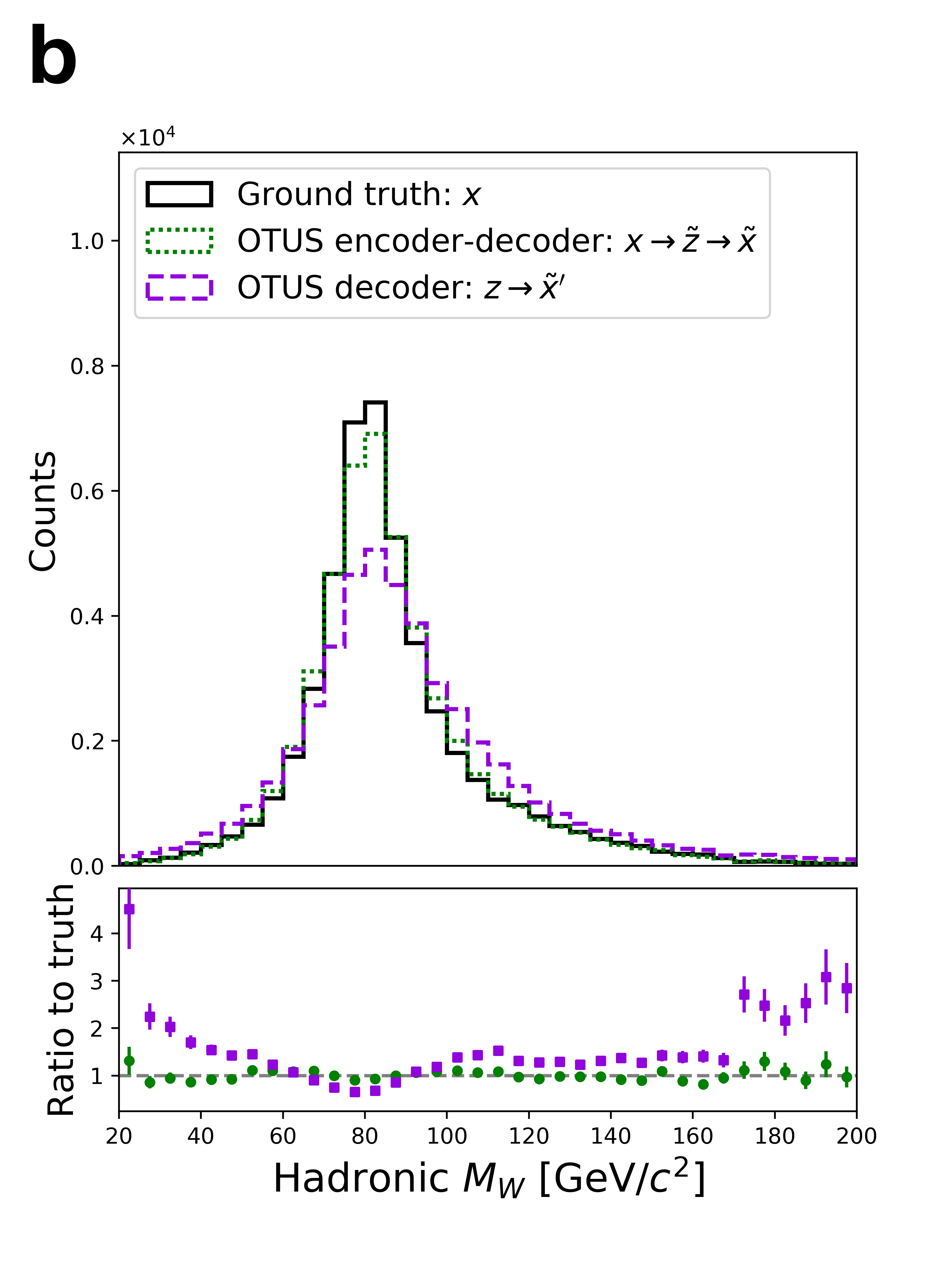}
    \end{subfigure}
    \begin{subfigure}{0.40\textwidth}
    \includegraphics[width=\linewidth]{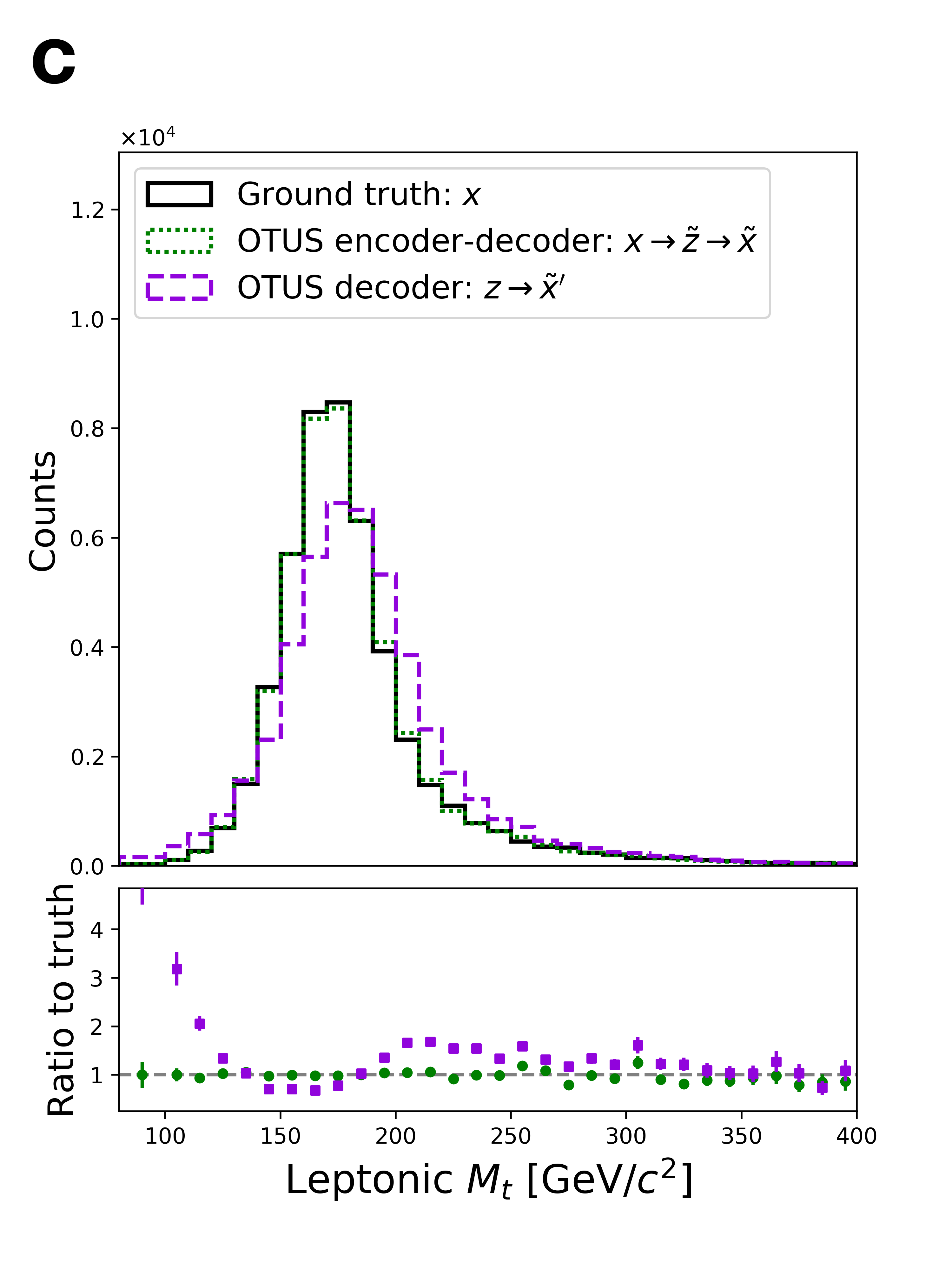}
    \end{subfigure}
    \begin{subfigure}{0.40\textwidth}
    \includegraphics[width=\linewidth]{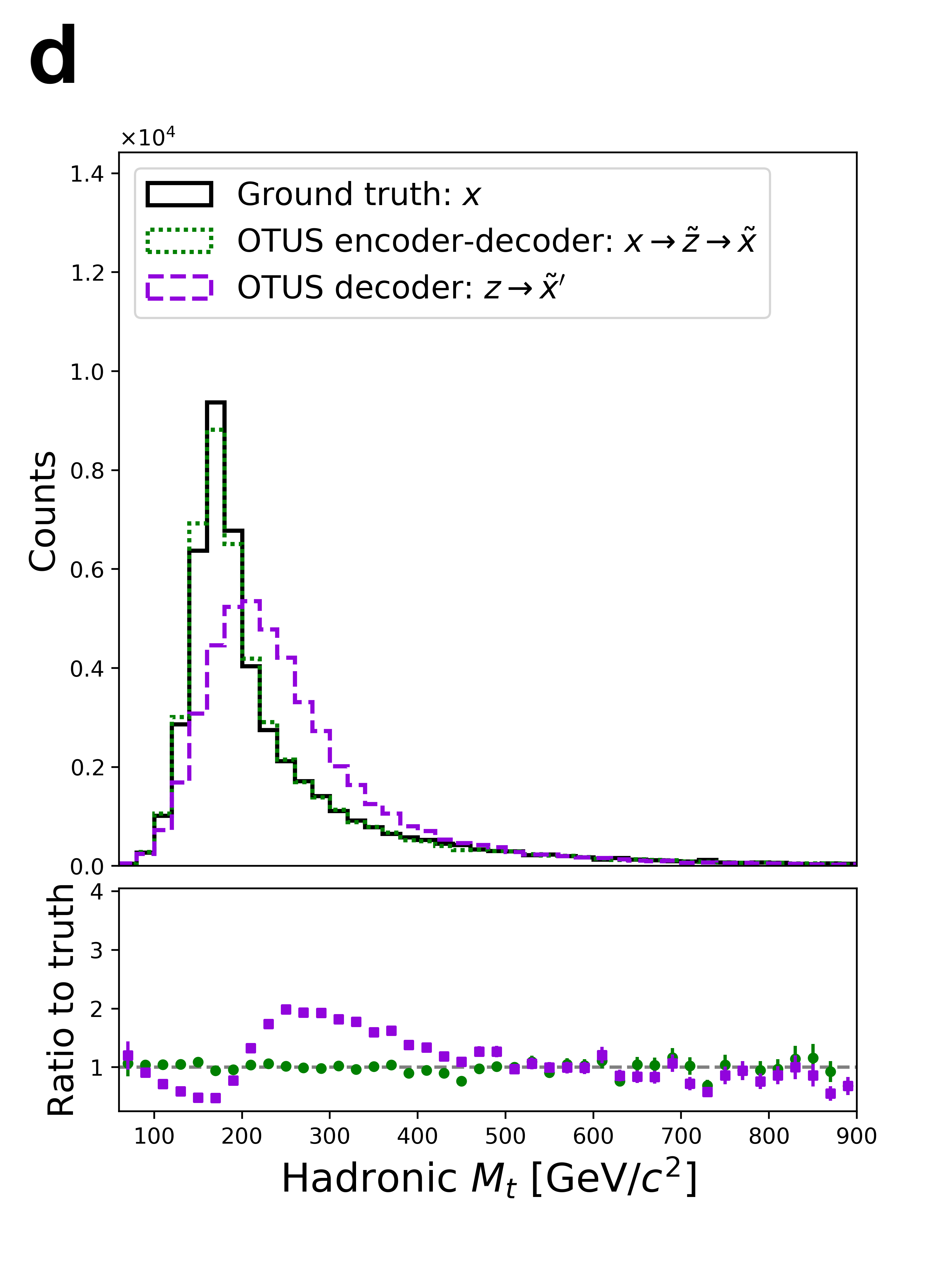}
    \end{subfigure}
    \caption{\E{\textbf{Performance of OTUS for semileptonic $t\bar{t}$ decays in  physically important derived quantities in $\mathcal{X}$.}}
    \E{\textbf{a} Matching of the invariant mass of the combined $t\bar{t}$ pair.}
    \E{\textbf{b} Matching of the invariant mass of the hadronically decaying \E{$W$-boson}, $M_W$.}
    \E{\textbf{c} Matching of the invariant mass of the \E{top-quark}, $M_t$, reconstructed using information from the leptonically decaying \E{$W$-boson}.}
    \E{\textbf{d} Matching of the invariant mass of the \E{top-quark}, $M_t$, reconstructed using information from the hadronically decaying \E{$W$-boson}.}
    \E{These show the testing sample $\{x \sim p(x)\}$ (solid black) in the experimental space, $\mathcal{X}$, as well as output from the decoder applied to samples drawn from $p(z)$, labeled as $z\rightarrow \tilde{x}'$ (dashed purple). Also shown are samples passed through both the decoder and encoder chain, $x\rightarrow\tilde{z}\rightarrow\tilde{x}$ (dotted green). Dotted green and solid black distributions are matched explicitly during training. Enhanced differences between dashed purple and solid black indicate the encoder's output needs improvement, as $p_E(z)$ does not fully match $p(z)$.}
    Residual plots show bin-by-bin ratios with statistical uncertainties propagated accordingly (see \nameref{subsec:eval}).}  
    \label{fig:t_mass}
\end{figure}

Finally, we examine the distribution of physically important derived quantities, the invariant masses of the \E{top-quarks} and \E{$W$-bosons} estimated by combining information from pairs and triplets of objects, see Figure~\ref{fig:t_mass}. No exact assignments are possible due to the ambiguity of the jet assignment and the lack of transverse information for the neutrino, but a comparison can be made between the experimental sample in $\mathcal{X}$ and the mapped samples $\mathcal{Z} \rightarrow \mathcal{X}$. As in the $Z \rightarrow e^+ e^-$ case, we see imperfect but reasonable matching on such derived quantities which the network was not explicitly instructed to learn.

\section{\CC{Methodology}} \label{sec:methods}
\CC{This section provides details on the methods used to produce the results in the previous section. We first describe the data generation process. We then describe the machine learning models used and strategies for how they were trained. Finally, we give details on the qualitative and quantitative evaluation methods used in the visualizations of the results.}

\subsection{Data Generation} \label{subsec:datagen}
The data for this work was generated with the programs Madgraph5 v.2.6.3.2~\cite{madgraph}, Pythia v.8.240~\cite{pythia}, and Delphes v.3.4.1~\cite{delphes}. ROOT v.6.08/00~\cite{root} was used to interface with the resulting Delphes output files. \C{We used the default run cards for Pythia, Delphes, and Madgraph. Where relevant, jets were clustered using the anti-kt algoritm~\cite{antiKtAlgorithm} with a jet radius of 0.5. The card files can be found with the code for this analysis (see \nameref{sec:codeavail}).} 

Samples of the physical latent space, $\mathcal{Z}$, were extracted from the Madgraph LHE files to form the 4-momenta of the particles. Samples of the data space, $\mathcal{X}$, were extracted from Delphes' output ROOT files. We selected for the appropriate final state: $e^+$, $e^-$ in the $Z \rightarrow e^+ e^-$ study and $e^-$, missing 4-momentum (i.e. $\E{{\rm MET}} = (\textbf{p}^\E{{\rm miss}},E^\E{{\rm miss}})$), and 4 jets in the semileptonic $t \bar{t}$ study. If an event failed this selection, the corresponding $\mathcal{Z}$ event was also  removed. Reconstructed data in $\mathcal{X}$ was extracted by default as $(p_\textrm{T}, \eta, \phi)$ of the object and converted into $(\textbf{p},E)$ via the following relations

\begin{align}
    \textbf{p} := (p_x, p_y, p_z) =& (p_\textrm{T} \operatorname{cos}(\phi), ~p_\textrm{T} \operatorname{sin}(\phi), ~p_\textrm{T} \operatorname{sinh}(\eta))\\
    E =& \sqrt{(p_\textrm{T} \operatorname{cosh}(\eta))^2 + m^2 },
\end{align}

\noindent where $m$ is the particle's definite mass and is zero for massless particles. Note that we are assuming natural units where the speed of light, $c$, is equal to unity. This equates the units of energy, $E$, momentum, $\textbf{p}$, and mass, $m$. In our case, $m_{e^+} = m_{e^-} = 0$ \E{[GeV${\rm c^{-2}}$]} is a standard assumption given that the true value is very small compared to the considered energy scales. We additionally set $m=0$ \E{[GeV${\rm c^{-2}}$]} for the 4 jets and $\E{{\rm MET}}$ since these objects have atypical definitions of mass. 

In total, we generated $491,699$  events for $Z \rightarrow e^+ e^-$ and $422,761$ events for semileptonic $t \bar{t}$. The last $160,000$ events in each case were  reserved solely for statistical tests after training and validation of OTUS.

\subsection{Model} \label{subsec:model}

\subsubsection{Model Choice} \label{subsec:modelchoice}
\C{
In this section, we briefly survey the literature of machine learning methods which might be considered for this task. We discuss their features and whether they are compatible choices for this application.

We will primarily focus on OT-based probabilistic autoencoder methods (i.e. WAE~\cite{wae} and its derivatives) but first we briefly address a derivative of VAEs, $\beta$-VAE. This method appears similar to WAE in the form of loss function that is used. Both have a data-space loss and a latent-space loss with a relative hyperparameter weighting $\beta$ (or $\lambda$ for the WAE). However, the $\beta$-VAE method is not principled in OT and thus is distinct from the WAE method and its derivatives. Most importantly for our application, the $\beta$-VAE (like its predecessor VAE) is likelihood-based which precludes it from applications where the latent prior is not analytically known. The interested reader can find more information on these distinctions in the following reference~\cite{vegan-cookbook}.

The WAE method~\cite{wae} provides a general framework for an autoencoder whose training is based on ideas from OT theory, namely the Wasserstein distance. This work defined a large umbrella under which a rich amount of subsequent literature falls (e.g. SWAE~\cite{swae}, Sinkhorn Autoencoders~\cite{sae}, CWAE~\cite{cwae}). The key difference between these methods and the original WAE method is the fact that each chooses a different $d_z(\cdot, \cdot)$ cost function. Therefore, the choice of method largely comes down to finding a suitable $d_z$ for the given problem.

The original WAE work proposes two specific options for the $d_z$, defining two versions of WAE: GAN-WAE and MMD-WAE. The first is an adversarial approach in which $d_z$ is the Jensen-Shannon divergence estimated using a discriminator network. The second chooses $d_z$ to be the Maximum Mean Discrepancy (MMD)~\cite{wae}.

The GAN-WAE strategy suffers from the same practical issues as other adversarial methods such as GANs (i.e. mode collapse). This possibility of training instability makes it an undesirable choice. The MMD-WAE does not have this training instability issue but requires an a priori choice of a kernel for the form of latent space prior, $p(z)$. This implies that we analytically know the desired prior form ahead of time, which is not the case for particle physics in general. Therefore, this option will not work for the applications explored in this work.

We now explore WAE derivatives which choose other choices for $d_z$ that might be more amenable to our application. CWAE~\cite{cwae} chooses the Cramer-Wold distance as the $d_z$ cost function. For a Gaussian latent space prior, this provides a computationally efficiency boost due to the existence of a closed-form solution. However, this assumption makes it unsuitable for our current application because our latent prior, $p(z)$, is non-Gaussian and often does not have a form which is known analytically a priori.

Two other derivatives allow for a flexible prior form which would be suitable for the task at hand. SWAE~\cite{swae} chooses the $d_z$ cost function to be the SW distance and Sinkhorn Autoencoder (SAE)~\cite{sae} chooses it to be the Sinkhorn divergence which is estimated via the Sinkhorn algorithm. Both have comparable performance with trade-offs in performance and computational efficiency. SAE claims superior performance to SWAEs for Gaussian priors, while it is slightly more computationally intensive ($\mathcal{O}(M^2)$ as opposed to SWAEs best case $\mathcal{O}(M)$ or worst case $\mathcal{O}(M {\rm log} M)$). However, both methods are valid choices for this application. Therefore, we suggest that SAE performance on this task be explored in future work. 

We also note the existence of other WAE-derivative methods which generalize the underlying OT framework. In our application, the $d_z$ metric always compares distributions in the same ambient space $\mathcal{Z}$. Additionally, the overall loss function also approximates the Wasserstein distance between two distributions in the same ambient space $\mathcal{X}$, namely $W_c(p(x), p_D(x))$. However, recent work using the Gromov-Wasserstein distance~\cite{vayer2020sliced} extends the underlying Optimal Transport (OT) framework to situations where the two probability measures $\mu$ and $\nu$ are not defined on the same ambient space (e.g. $\mathbb{R}^n$ and $\mathbb{R}^m$ with different dimensions $n$ and $m$). For this application, this is an over-powered tool since by construction $p(z)$ and $p_E(z)$ ($p(x)$ and $p_D(x)$) always lie in the same ambient space. However, if one were attempting to study the optimal transportation between different spaces, this would be ideal. This would be an interesting direction to follow-up recent related work which connects OT and particle physics~\cite{LOT, Komiske_2020}.
}

\subsubsection{Base Model} \label{subsubsec:genmodel}
Both the encoder and decoder models of OTUS are implicit conditional generative models, and operate by concatenating the input with random noise and passing the resulting vector through feedforward neural networks.

 For a model, $G$, mapping from a space, $\E{\mathcal{U}}$, to a space, $\E{\mathcal{V}}$, the steps are as follows. 
 (1) A sample of raw input data, $u \in \E{\mathcal{U}}$, is standardized by subtracting the mean and dividing by the standard deviation resulting in the standardized data \C{vector}, $\bar{u}$.
 (2) A noise neural network computes a conditional noise distribution $p_N(\epsilon  \mid  u)$, where the noise vector $\epsilon \sim p_N(\epsilon  \mid  u)$ has the same dimensionality as the core network prediction $\bar{w}$ (defined in the next step).
 (3) The standardized data \C{vector}, $\bar{u}$, and noise vector, $\epsilon$, are then concatenated and fed into a core neural network. This network outputs the 3-momentum, $\textbf{p}$, information of each particle in the standardized space, collected into a vector $\bar{w}$.
 (4) The vector $\bar{w}$ is then unstandardized by inverting the relationship in step 1, creating a vector $w$.
 (5) The Minkowski relation \CC{( see \nameref{subsec:approach})} is then enforced explicitly to reinsert the energy information of each particle, transforming $w$ into the final $v \in \E{\mathcal{V}}$ which is distributed according to $p_G(v \mid u)$.

 Both the encoding and decoding model's noise networks produce Gaussian-distributed noise vectors with mean and diagonal covariances $[ \mu(x), \sigma^2(x))]$ and $[ \mu(z), \sigma^2(z))]$ respectively. 
For the $Z \rightarrow e^+ e^-$ study, the core and noise networks for both the encoder and decoder each used a simple feed-forward neural network architecture with a single hidden layer, with 128 hidden units and ReLU activation.

\subsubsection{Model for semileptonic \E{top-quark} decay study} \label{subsubsec:ttbarmodel}
To better model the complexities in the semileptonic $t\bar{t}$ data, we introduced a restriction to the decoder model and modified the training procedure accordingly (see \nameref{subsec:training}). With these modifications, the base model encountered difficulty during training, so we introduced the following three changes to the architecture for more effective training.

First, 
the conditionality of the noise network is removed and the noise 
is instead drawn from a fixed standard normal distribution, $p_N(\epsilon \mid u) = p_N(\epsilon) = \mathcal{N}(\mathbf{0}, \mathbf{I})$. Second, the model now has a residual connection such that the core network now predicts the change from the input $u$. The 3-momentum sub-vector of $u$ is added to $w$ before proceeding to imposing the Minkowski relation in step 5. 
This input-to-output residual connection provides an architectural bias towards identity mapping, when the model is initialized with small random weights.

Lastly, the core network itself is augmented with residual connections~\cite{ResNet} and batch normalization~\cite{ioffe2015batchnorm}. An input vector to the core network is processed as follows: 
(A) A linear transform layer with $K$ units maps the input to a vector $r \in \mathbb{R}^K$. 
(B) Two series of \C{$[\text{BatchNorm, ReLU, Linear}]$} layers are applied sequentially to $r$, without changing the dimensionality, resulting in $s \in \mathbb{R}^K$. 
(C) A residual connection from $r$ is introduced, so that $s \rightarrow s + r$. 
(D) The resulting $s$ is then transformed by a final linear layer with $J$ units to obtain the output vector $t \in \mathbb{R}^J$. 
For the $t\bar{t}$ study, the input vector $[\bar{u}, \epsilon]$ is $24+18=42$ dimensional, the output dimension $J=18$, and we set $K=64$ for the core network, in both the encoder and decoder models.

\subsection{Training} \label{subsec:training}

\subsubsection{Base Training Strategy} \label{subsubsec:gentraining}
As described \CC{in \nameref{subsec:approach},} the model is trained by minimizing the SWAE loss function augmented with anchor terms

\begin{align}
\begin{aligned}
\mathcal{L}_\E{{\rm SWAE}}(p(x), p_D(x \mid z), p_E(z \mid x)) =& ~ \mathbb{E}_{x \sim p(x)} \mathbb{E}_{z \sim p_E(z \mid x)} \mathbb{E}_{\tilde{x} \sim p_D(x \mid z)} [c(x,  \tilde{x})]  + \lambda \E{d_{\rm SW}}(p_E(z), p(z)) \\
& + \beta_E \mathcal{L}_A(p(x), p_E(z \mid x)) + \beta_D \mathcal{L}_A(p(z), p_D(x \mid z)),  \label{eq:overall-loss}
\end{aligned}
\end{align}

\noindent with respect to parameters of the encoder $p_E(z \mid x)$ and decoder $p_D(x \mid z)$ distributions.

As each term in the loss function has the form of an expectation, we approximate each with samples and compute the following Monte-Carlo estimate of the loss:

\begin{align}
\begin{aligned}
\hat{\mathcal{L}}_\E{{\rm SWAE}} =& ~ \frac{1}{M} \sum_{m=1}^M c(x_m,  \tilde{x}_m) + \lambda \frac{1}{L * M} \sum_{l=1}^{L} \sum_{m=1}^{M} c((\theta_l \cdot z_m)_\E{{\rm sorted}}, (\theta_l \cdot \tilde{z}_m)_\E{{\rm sorted}}) \\
& + \beta_E \frac{1}{M} \sum_{m=1}^M c_A(x_m, \tilde{z}_m) + \beta_D \frac{1}{M} \sum_{m=1}^M c_A(z_m, \tilde{x}'_m),
\end{aligned}
\end{align}

\noindent where $\{x_m\}_{m=1}^M$ and $\{z_m\}_{m=1}^M$ are $M$ instances of $\mathcal{X}$ and $\mathcal{Z}$ samples, $\{\tilde{z}_m \sim p_E(\cdot  \mid  x_m) \}_{m=1}^M$ are drawn from the encoder, $\{\tilde{x}'_m \sim p_D(\cdot  \mid  z_m) \}_{m=1}^M$ are drawn from the decoder, and $\{\tilde{x}_m \sim p_D(\cdot  \mid  \tilde{z}_m) \}_{m=1}^M$ are drawn from the auto-encoding chain $x \to \tilde{z} \to \tilde{x} $.%
\footnote{This is equivalent to drawing a sample $(x, \tilde{z}, \tilde{x})$ from the joint distribution $p(x)p_E(\tilde{z} \mid x) p_D(\tilde{x} \mid \tilde{z})$.} The estimation of  $\E{d_{\rm SW}}(p(z), p_E(z))$ uses $L$ random slicing directions $\{\theta_l\}_{l=1}^L$ drawn uniformly from the unit sphere, along which the samples $z_m \sim p(z)$ and $\tilde{z}_m \sim p_E(z)$ are compared; this involves estimating each $\E{{\rm CDF}}^{-1}$ by sorting the two sets of projections in ascending order as $\{(\theta_l \cdot z_m)_\E{{\rm sorted}}\}_{m=1}^M$ and $\{(\theta_l \cdot \tilde{z}_m)_\E{{\rm sorted}}\}_{m=1}^M$, for each direction $\theta_l$; we refer interested readers to \cite{swae} for more technical details of the Sliced Wasserstein distance. We use the squared norm as the cost metric $c(u,v)=||u-v||^2$ in the SWAE loss \cite{swae}. The anchor cost, $c_A$, between two observation vectors $u, v$ (which can reside in either $\mathcal{X}$ or $\mathcal{Z}$ space) is defined as $c_A(u, v):= 1 - \hat{\textbf{p}}_u \cdot \hat{\textbf{p}}_v$, where $\hat{\textbf{p}}_u$ is the unit vector of the coordinates of $u$ corresponding to the momentum of a pre-specified particle, and $\hat{\textbf{p}}_v$ is defined analogously with respect to the same particle; this is chosen as the electron in our experiments.
For example, $c_A(x, \tilde{z})$ would be computed as 

\begin{align}
    c_A(x, \tilde{z}) = 1 -  \hat{\textbf{p}}^{e-}_x \cdot \hat{\textbf{p}}^{e-}_{\tilde{z}} = 1 - \frac{\textbf{p}^{e-}_x}{\|\textbf{p}^{e-}_x\|} \cdot \frac{\textbf{p}^{e-}_{\tilde{z} }} {\|\textbf{p}^{e-}_{\tilde {z}}\|}.
\end{align}

\noindent At a higher level, the computation of $\hat{\mathcal{L}}_\E{{\rm SWAE}}$ based on a mini-batch proceeds as follows. 
Following the path through the full model, a batch of samples $X \sim p(x)$ from $\mathcal{X}$ space is passed to the encoder model, $E$, producing $\tilde{Z} \in \mathcal{Z}$ distributed according to $p_E(z \mid x)$. The encoding anchor loss term $L_{A,E}(X, \tilde{Z}) \equiv \mathcal{L}_A (p(x), p_E(z \mid x))$ is then computed along with the SW distance latent loss, $\hat{d}_\E{{\rm SW}}(Z, \tilde{Z}) \equiv  \hat{d}_\E{{\rm SW}}(p(z), p_E(z))$. The samples $\tilde{Z}$ and $Z \sim p(z)$ are then passed independently in parallel through the decoder model, $D$, producing $\tilde{X}$ and $\tilde{X}'$, respectively. The decoding anchor loss term $L_{A,D}(Z, \tilde{X}') \equiv \mathcal{L}_A (p(z), p_D(x \mid z))$ is then computed. Finally, the data space loss, chosen to be $\E{{\rm MSE}}(X, \tilde{X})$, is computed. 
See \EE{Supplementary Figure 1} for a visual representation. 
We can then minimize the tractable Monte-Carlo estimate of the objective, $\hat{\mathcal{L}}_\E{{\rm SWAE}}$, by stochastic gradient descent with respect to parameters of the encoder and decoder networks.

Since the original (S)WAE aimed to ultimately minimize $d_W(p(x), p_D(x))$ via an approximate variational formulation,%
\footnote{When minimized over all $p_E(z \mid x)$ that satisfies the constraint $p_E(z) = p(z)$, term A of \E{Equation} (\ref{eq:wae}) becomes an upper bound on $d_{W}(p(x), p_D(x))$; the bound is tight for deterministic decoders \cite{wae}. The overall WAE loss \E{$\mathcal{L}_{\rm WAE}$} is a relaxation of the exact variational bound, and recovers the latter as $\lambda \to \infty$.} we also consider an auxiliary strategy of directly minimizing the more computationally convenient SW distance $\E{d_{\rm SW}}(p(x), p_D(x))$ to train a decoder, or minimizing $\E{d_{\rm SW}}(p(z), p_E(z))$ to train an encoder. This can be done by simply optimizing the Monte-Carlo estimates

\begin{align}
    \hat{d}_\E{{\rm SW}}(p(x), p_D(x)) = \frac{1}{L * M} \sum_{l=1}^{L} \sum_{m=1}^{M} c((\theta_l \cdot x_m)_\E{{\rm sorted}}, (\theta_l \cdot \tilde{x}_m')_\E{{\rm sorted}})  \\
    \hat{d}_\E{{\rm SW}}(p(z), p_E(z)) = \frac{1}{L * M} \sum_{l=1}^{L} \sum_{m=1}^{M} c((\theta_l \cdot z_m)_\E{{\rm sorted}}, (\theta_l \cdot \tilde{z}_m)_\E{{\rm sorted}}), 
\end{align}

\noindent where the samples $\{x_m, z_m, \tilde{x}'_m, \tilde{z}_m\}_{m=1}^M$ are defined the same way as before.
Auxiliary training of the encoder was found helpful for escaping local minima when optimizing the joint loss $\hat{\mathcal{L}}_\E{{\rm SWAE}}$, and auxiliary fine-tuning of the decoder in post-processing also improved the decoder's fit to the data. 

Note that the idea of training a decoder by itself is similar in spirit to GANs, but again with the major distinction and innovation that we use samples from a physically meaningful prior $p(z)$ instead of an uninformed generic one (e.g. Gaussian), as we are also interested in a physical conditional mapping $p_D(x \mid z)$ in addition to achieving good fit to the marginal $p(x)$. 

\subsubsection{Training an (S)WAE with a restricted decoder} \label{subsubsec:restrictedtrain}
As was previously explained, experimental limitations in the semileptonic $t\bar{t}$ study require a minimum threshold, so that jets which have $p_{\textrm{T}} < 20$ \E{[GeV${\rm c^{-1}}$]} are discarded and therefore  are not represented in the training dataset, as they would not be available in control region data. Denoting the region of $\mathcal{X}$ space which passes this threshold by $S$, we are faced with the task of fitting a distribution $p_D(x)$ over $\mathcal{X}$ while only having access to data samples in the valid subset $S \subset \mathcal{X}$.

We propose a general method for fitting an (S)WAE such that its marginal data distribution $p_D(x)$, when restricted to the valid set $S$, matches that of the available data. We first define the \E{restricted} marginal data distribution,

\begin{align}
    \bar{p}_D(x) = \frac{p_D(x) \mathbf{1}_S(x)}{\C{P_D(S)}},
\end{align}

\noindent where  $\mathbf{1}_S(x)$ is the indicator function of $S$ so that it equals $1$ if $x  \in S$, and $0$ otherwise, and $ \C{P_D(S)} := \int dt p_D(t) \mathbf{1}_S(t)$ normalizes this distribution. Note that $\C{P_D(S)}$ depends on the decoder parameters, and can be identified as the probability that the data model $p_D(x)$ yields a valid sample $x \in S$. 

Our goal is then to minimize $d_W(p(x), \bar{p}_D(x))$. This can be done by minimizing the same variational upper bound as in a typical (S)WAE, but with an adjustment to the data loss function in term A of \E{Equation} (\ref{eq:wae}), so it becomes

\begin{align}
\mathbb{E}_{x \sim p(x)} \mathbb{E}_{p_E(z \mid x)} \mathbb{E}_{\tilde{x} \sim \bar{p}_D(x \mid z)} [c(x,  \tilde{x})] \rightarrow
\mathbb{E}_{x \sim p(x)} \mathbb{E}_{p_E(z \mid x)} \mathbb{E}_{\tilde{x} \sim p_D(x \mid z)} [\frac{\mathbf{1}_S(\tilde{x})}{\C{P_D(S)}}  c(x,  \tilde{x})] \label{eq:restricted_dataloss}.
\end{align}

\noindent Letting $\theta$ denote the parameters of the model, it can be shown that the gradient of the modified cost function has the simple form

\begin{align}
\nabla_\theta \frac{\mathbf{1}_S(\tilde{x})} {\C{P_D(S)}} c(x,  \tilde{x})  = \frac{\mathbf{1}_S(\tilde{x})} {\C{P_D(S)}}  \nabla_\theta c(x,  \tilde{x}).
\end{align}

\noindent This means that training an (S)WAE with a restricted decoder by stochastic gradient descent proceeds as in the unrestricted base training strategy, except that only the valid samples in $S$ contribute to the gradient of the data loss term, with the contribution scaled inversely by the factor $\C{P_D(S)}$, which can be estimated by drawing samples $\tilde{x}_m' \sim p_D(x)$%
\footnote{This is equivalent to passing $z_m \sim p(z)$ through the decoder to produce $\tilde{x}_m'$.} and forming the  Monte-Carlo estimate

\begin{align}
\C{P_D(S)} \approx \frac{1}{M} \sum_{m=1}^M \mathbf{1}_S(\tilde{x}'_m).
\end{align}

\subsubsection{Parameter Optimization} \label{subsubsec:trainparams}
For the $Z \rightarrow e^+ e^-$ study, we used the base training strategy. We optimized $\hat{\mathcal{L}}_\E{{\rm SWAE}}$ for 80 epochs with anchor penalties $\beta_E = \beta_D = 50$, followed by another 800 epochs with the anchor penalties set to 0.
For the semileptonic $t \bar{t}$ study,
we modified the base training strategy to accommodate a restricted decoder, substituting all appearances of $p_D(x)$ in the loss $\hat{\mathcal{L}}_\E{{\rm SWAE}}$ by $\bar{p}_D(x)$ (e.g. using the modified data loss term \E{Equation} (\ref{eq:restricted_dataloss})). 
We optimized the resulting loss $\hat{\mathcal{L}}_\E{{\rm SWAE}}$ till convergence, for about 1000 epochs. Then we froze the encoder and  fine-tuned the decoder by minimizing $\hat{d}_\E{{\rm SW}}(p(x), \bar{p}_D(x))$ for 10 epochs, with a reduced learning rate.
The input-to-output residual connection (see \nameref{subsec:model}) in the $t\bar{t}$ model allowed for sufficiently high $\C{P_D(S)} \approx 0.6$ and reliable gradient estimates during training, and the architectural bias towards identity mapping made the anchor losses redundant,  so we set $\beta_E = \beta_D = 0$.

In both studies, we found that a sufficiently large batch size significantly improved results. This is likely do to increasing the accuracy of gradient estimates for stochastic gradient descent and also the $\E{{\rm CDF}}^{-1}$ in the SWAE latent loss. In all of our experiments, we used the Adam optimizer~\cite{kingma2015adam} with $L=1,000$ number of slices, a batch size of $M = 20,000$, and learning rate of 0.001. 
We tuned the $\lambda$ hyperparameter of the (S)WAE loss $\hat{\mathcal{L}}_\E{{\rm SWAE}}$ on the validation set; we set $\lambda=1$ for the $Z \rightarrow e^+ e^-$ model, and $\lambda=20$ for the $t \bar t$ model.

\subsection{Evaluation} \label{subsec:eval}
\CC{
This section provides details on the various qualitative and quantitative evaluation techniques used in this work.
}

\CC{
As common in the literature~\cite{calogan, ganlhc, lagan}, we visualize our results along  informative one-dimensional projections using histograms (e.g. Figure \ref{fig:z_space} and Figure \ref{fig:z_mass}). 
We choose the bin sizes such that the error on the counts can be approximated as Gaussian distributed.
These histograms are accompanied by residual plots, showing the ratio between the histograms from generated samples and the histogram from true samples, with accompanying statistical errors~\cite{statsRef}%
\footnote{Specifically, for a bin with counts $h_{1}$ and $h_{2}$, respectively, the error on the ratio, $r = h_{2}/h_{1}$ is $\sigma_{r} = r \sqrt{\frac{1}{h_{2}} + \frac{1}{h_{1}}}$.}. We also visualize the generative mappings using transportation plots (e.g. Figure \ref{fig:z_trans}) that 
allow us to confirm the physicality of the learned mappings.
]}

\CC{
In addition to qualitative comparisons, we also evaluated the results using several quantitative metrics. To this end, we calculate the Monte-Carlo estimate of the SW distance, $\hat{d}_{\rm SW}(\cdot, \cdot)$, using $L=1,000$ slices according to the cost metric $c(u,v) = || u - v ||^2$. The results are reported for each study in the text. In addition, we 
 apply several statistical tests on the considered one-dimensional projections, which we report in Supplementary Tables 1-4. 
First, we calculate the reduced $\chi^2$, $\chi^2_R$, for each comparison and report it along with the degrees-of-freedom (dof). 
Second, we calculate the unbinned two-sample, two-sided Kolmogorov-Smirnov distance. 
Lastly, we calculate the Monte-Carlo estimate of the Wasserstein distance, $\hat{d}_{W}(\cdot, \cdot)$, according to the cost metric $c(u,v) = || u - v ||^2$. 
All statistical tests were carried-out using two separate test sets not used during training or validation of the networks. 
The number of samples in each test set were $80,000$ in the $Z \rightarrow e+ e-$ study and $47,856$ in the semileptonic $t \bar{t}$ study.%
\footnote{Note that the number of samples in the semileptonic $t \bar{t}$ study is lower due to the hard $p_{\rm T}$ cutoff constraint as described in section \ref{subsec:demottbar}.The events present are ones that passed this cutoff constraint.}
}

\section{\CC{Conclusion}} \label{sec:discussion}

OTUS is a data-driven, machine-learned\E{,} predictive simulation strategy which suggests a possible new direction for alleviating the prohibitive computational costs of current Monte-Carlo approaches, while avoiding the inherent disadvantages of other machine-learned approaches.  We anticipate that the same ideas can be applied broadly outside of the field of particle physics.

In general, OTUS can be applied to any process where unobserved latent phenomena $\mathcal{Z}$ can be described in the form of a prior model, $p(z)$, and are translated to an empirical set of experimental data, $\mathcal{X}$, via an unknown transformation.   For example, in molecular simulations in \E{chemistry} observations could be measurements of real-world molecular dynamics, $p(z)$ would represent the model description of the system, and $p(x \mid z)$ would model the effects of real-world complications~\cite{otherfields_chem}. In \E{cosmology}, $\mathcal{X}$ could be the distribution of mass in the observed universe, $p(z)$ could describe its distribution in the early universe, and $p(x \mid z)$ would model the universe's unknown expansion dynamics (e.g. due to inflation)~\cite{Seljak_2017, otherfields_cosmology}. In \E{climate simulations}, $p(z)$ could correspond to the climate due to a physical model, while $p(x \mid z)$ takes unknown geography-specific effects into account~\cite{otherfields_climate}. Additionally, an immediate and promising application of OTUS is in \E{medical imaging}, which uses particle physics simulations to model how the imaging particles (e.g x-rays) interact with human tissue and suffers from the great computational cost of these simulations~\cite{otherfields_photonics}. 
\CC{We note that our method assumes a high degree of mutual information between $\mathcal{Z}$ and $\mathcal{X}$ in the desired application. Therefore, in situations where such mutual information is low (e.g. chaotic turbulent flows) the transformations learned by this method would likely be less reliable.}

Moreover, features of this method can be \E{adapted} to suit the particular problem's needs. For example, in this work we were interested in low-dimensional data, however the method could also be applied to high-dimensional datasets.  Moreover, the encoding and decoding mappings can be stochastic, as in this work, or deterministic. Lastly, while this work aimed to be completely unsupervised, and thus data-driven, OTUS can be easily extended to a semi-supervised setting. In this case, the data would consist mostly of unpaired samples but would have a limited number of paired examples \C{$(z, x)$} (e.g. from simulation runs). 
These pairs sample the joint distribution, $p(z, x)$, which, combined with the decoder $p_D(\tilde{x} \mid z)$, yields a transportation map $\gamma$ between $p(x)$ and $p_D(\tilde{x})$, $\gamma(p(x), p_D(\tilde{x})) :=\int dz p(z,x) p_D(\tilde{x} \mid z)$. Since calculating the Wasserstein distance between $p(x)$ and $p_D(\tilde{x})$ involves finding the optimal transportation map, this particular choice yields an upper bound on the Wasserstein distance. We can similarly construct a transportation map between $p(z)$ and $p_E(\tilde{z})$ using $p(z,x)$ and $p_E(z \mid x)$. This makes directly optimizing the Wasserstein distances $d_W(p(x), p_D(x))$ and $d_W(p(z), p_E(z))$ tractable in this high-dimensional setting. Therefore, we get the alternative objectives

\begin{align}
    \mathcal{L}_\E{{\rm paired}}(p_D(x \mid z), p(z,x)) = \mathbb{E}_{(z,x) \sim p(z,x)} \mathbb{E}_{\tilde{x} \sim p_D(x \mid z)} [c(x, \tilde{x})] \label{eq:semiSupW_data}\\
    \mathcal{L}_\E{{\rm paired}}(p_E(z \mid x), p(z,x)) = \mathbb{E}_{(z,x) \sim p(z,x)} \mathbb{E}_{\tilde{z} \sim p_E(z \mid x)} [c(z, \tilde{z})] \label{eq:semiSupW_latent},
\end{align}

\noindent which are upper bounds on $d_W(p(x), p_D(x))$ and $d_W(p(z), p_E(z))$ respectively. These terms can be incorporated alongside the unsupervised SWAE loss, to leverage paired examples \C{$\{(z, x) \sim p(z,x)\}$} in a semi-supervised setting.

We have demonstrated the ability of OTUS to learn a detector transformation in an unsupervised way. The results, while promising for this initial study, leave room for improvement. Several directions could lead to higher fidelity descriptions of the data and latent spaces.

First, the structure of the latent and data spaces can significantly affect the performance and \CC{physicality} of the resulting simulations. Particle physics data has rich structures often governed by group symmetries and conservation laws. Our current vector format description of the data omits much of this complicated structure. For example, we omitted categorical characteristics of particles like charge and type. Knowledge of such properties and the associated rules likely would have excluded the necessity of terms like the anchor loss.  Therefore, future work should explore network architectures and losses that can better capture the full nature of these data structures~\cite{GNsym, NNsym}.
 
The next technical hurdle is the ability to handle variable input and output states. The same $p(z)$ can lead to different detected states as was described, but not explored, in the semileptonic $t\bar{t}$ study where the number of jets can vary. Additionally, it should be possible to handle mixtures of underlying priors in the latent space. This can cause the number and types of latent-space particles to vary from one sample to another. For example, the Z boson can decay into $Z \rightarrow \mu^+ \mu^-$ in addition to $Z \rightarrow e^+ e^-$; a simulator should be able to describe these two cases holistically.
  
Finally, an essential feature of a predictive simulator is that it learns a general transformation, allowing it to make predictions for points in the latent space which lie outside of the control regions. This would require structuring the latent and data spaces to accommodates data from several control regions, such that the network may learn to interpolate between them.  Since networks excel at interpolation we expect that this will be a straightforward step.

\section{Data Availability} \label{sec:dataavail}
The datasets generated and analysed during the current study are available in the DRYAD repository, doi: \href{https://doi.org/10.7280/D1WQ3R}{10.7280/D1WQ3R}. 

\section{Code Availability} \label{sec:codeavail}
The code used during the current study are available in the Zenodo repository and are linked to the dataset, doi: \href{https://doi.org/10.5281/zenodo.4706055}{10.5281/zenodo.4706055}.



\section{\C{Acknowledgments}}
\E{J.N.H.} acknowledges support by the National Science Foundation under grants DGE-1633631 and DGE-1839285, and U.S. Department of Energy, Office of Science under the grant DE-SC0009920.
\E{Y.Y.} acknowledges funding from the Hasso Plattner Foundation.
\E{S.M.} acknowledges support by the National Science Foundation under Grants 2047418, 1928718,
2003237 and 2007719, Intel, Disney, Qualcomm, the U.S. Department of Energy, Office of Science under the grant SC0022331, and the Defense Advanced Research Projects Agency (DARPA) under Contract No. HR001120C0021. Any opinions,
findings and conclusions or recommendations expressed in this material are those of the author(s) and
do not necessarily reflect the views of the Defense Advanced Research Projects Agency (DARPA) or the National Science Foundation.

\section{Author Contributions Statement}
Using the CASRAI CRediT Contributor Roles Taxonomy: Conceptualization, \E{J.N.H.}, \E{S.M.}, \E{D.W.}, \E{Y.Y.}; Data curation, \E{J.N.H.}; Formal analysis, \E{J.N.H.}, \E{Y.Y.}; Funding acquisition, \E{J.N.H.}, \E{S.M.}, \E{D.W.}; Investigation, \E{J.N.H.}, \E{S.M.}, \E{D.W.}, \E{Y.Y.}; Methodology, \E{J.N.H.}, \E{Y.Y.}; Project administration, \E{J.N.H.}, \E{S.M.}, \E{D.W.}; Software, \E{J.N.H.}, \E{Y.Y.}; Supervision, \E{S.M.}, \E{D.W.}; Validation, \E{Y.Y.}; Visualization, \E{J.N.H.}; Writing – original draft, \E{J.N.H.}, \E{D.W.}; Writing – review \& editing, \E{J.N.H.}, \E{S.M.}, \E{D.W.}, \E{Y.Y.}.

\section*{Competing Interests Statement}
The authors declare no competing interests.
%

\section{Supplementary Information}  \label{sec:suppInfo}
\beginsupplement

\subsection{Supplementary Statistics}
\begin{table}[h]
\centering
\label{ta:ppZee_zspace}
\begin{tabular}{|p{0.8in}|p{0.8in}|p{0.8in}|p{0.8in}|}
\hline
&\multicolumn{3}{|c|}{$z$ vs $\tilde{z}$}\\\cline{2-4}
& \footnotesize{\textbf{W [GeV$^2$]}} & \footnotesize{\textbf{($\chi^2_R$, dof)}} & \footnotesize{\textbf{KS}}\\\hline \hline
\footnotesize{\textbf{Figure 3a ($p_y$)}} & \footnotesize{$1.34 \times 10^{+00}$} & \footnotesize{($50.583$, $23$)} & \footnotesize{$1.61 \times 10^{-02}$}\\\cline{2-4}
\footnotesize{\textbf{Figure 3a ($p_z$)}} & \footnotesize{$1.59 \times 10^{+00}$} & \footnotesize{($1.325$, $26$)} & \footnotesize{$4.90 \times 10^{-03}$}\\\cline{2-4}
\footnotesize{\textbf{Figure 3a ($E$)}} & \footnotesize{$1.29 \times 10^{+00}$} & \footnotesize{($8.814$, $26$)} & \footnotesize{$1.47 \times 10^{-02}$}\\\hline
\footnotesize{\textbf{Figure 5a}} & \footnotesize{$2.73 \times 10^{+01}$} & \footnotesize{($822.762$, $39$)} & \footnotesize{$2.46 \times 10^{-01}$}\\\hline
\end{tabular}
\caption{Table showing $\mathcal{Z}$ space statistical test results for the $Z \rightarrow e^+e^-$ dataset. These tests were performed on the distributions in the referenced figures in the main text. ${\rm W}$ is the Wasserstein distance, $\chi_R^2$ is the reduced $\chi^2$ and ${\rm dof}$ is the degrees-of-freedom, and ${\rm KS}$ is the value of the Kolmogorov-Smirnov statistical test. See the Evaluation section in the main text for detailed information about the calculations of these statistics.}
\end{table}

\begin{table}[h]
\centering
\label{ta:ppZee_xspace}
\begin{tabular}{|p{0.8in}|p{0.8in}|p{0.8in}|p{0.8in}|p{0.8in}|p{0.8in}|p{0.8in}|}
\hline
&\multicolumn{3}{|c|}{$x$ vs $\tilde{x}$}&\multicolumn{3}{|c|}{$x$ vs $\tilde{x}'$}\\\cline{2-7}
& \footnotesize{\textbf{W [GeV$^2$]}} & \footnotesize{\textbf{($\chi^2_R$, dof)}} & \footnotesize{\textbf{KS}} & \footnotesize{\textbf{W [GeV$^2$]}} & \footnotesize{\textbf{($\chi^2_R$, dof)}} & \footnotesize{\textbf{KS}}\\\hline \hline
\footnotesize{\textbf{Figure 3b ($p_y$)}} & \footnotesize{$4.22 \times 10^{-01}$} & \footnotesize{($1.391$, $23$)} & \footnotesize{$3.48 \times 10^{-03}$} & \footnotesize{$1.05 \times 10^{+00}$} & \footnotesize{($37.560$, $23$)} & \footnotesize{$1.22 \times 10^{-02}$}\\\cline{2-7}
\footnotesize{\textbf{Figure 3b ($p_z$)}} & \footnotesize{$3.71 \times 10^{+00}$} & \footnotesize{($1.523$, $26$)} & \footnotesize{$1.03 \times 10^{-02}$} & \footnotesize{$9.53 \times 10^{+00}$} & \footnotesize{($4.775$, $26$)} & \footnotesize{$7.49 \times 10^{-03}$}\\\cline{2-7}
\footnotesize{\textbf{Figure 3b ($E$)}} & \footnotesize{$6.64 \times 10^{-01}$} & \footnotesize{($0.489$, $26$)} & \footnotesize{$3.19 \times 10^{-03}$} & \footnotesize{$3.64 \times 10^{+00}$} & \footnotesize{($9.370$, $26$)} & \footnotesize{$2.00 \times 10^{-02}$}\\\hline
\footnotesize{\textbf{Figure 5b}} & \footnotesize{$7.28 \times 10^{-01}$} & \footnotesize{($5.055$, $39$)} & \footnotesize{$2.61 \times 10^{-02}$} & \footnotesize{$7.15 \times 10^{-01}$} & \footnotesize{($12.821$, $39$)} & \footnotesize{$3.14 \times 10^{-02}$}\\\hline
\end{tabular}
\caption{Table showing $\mathcal{X}$ space statistical test results for the $Z \rightarrow e^+e^-$ dataset. These tests were performed on the distributions in the referenced figures in the main text. ${\rm W}$ is the Wasserstein distance, $\chi_R^2$ is the reduced $\chi^2$ and ${\rm dof}$ is the degrees-of-freedom, and ${\rm KS}$ is the value of the Kolmogorov-Smirnov statistical test. See the Evaluation section in the main text for detailed information about the calculations of these statistics.}
\end{table}

\begin{table}[h]
\centering
\label{ta:ttbar_zspace}
\begin{tabular}{|p{0.8in}|p{0.8in}|p{0.8in}|p{0.8in}|}
\hline
&\multicolumn{3}{|c|}{$z$ vs $\tilde{z}$}\\\cline{2-4}
& \footnotesize{\textbf{W [GeV$^2$]}} & \footnotesize{\textbf{($\chi^2_R$, dof)}} & \footnotesize{\textbf{KS}}\\\hline \hline
\footnotesize{\textbf{Figure 6a ($p_y$)}} & \footnotesize{$1.58 \times 10^{+01}$} & \footnotesize{($7.418$, $49$)} & \footnotesize{$1.25 \times 10^{-02}$}\\\cline{2-4}
\footnotesize{\textbf{Figure 6a ($p_z$)}} & \footnotesize{$5.52 \times 10^{+01}$} & \footnotesize{($4.613$, $55$)} & \footnotesize{$1.65 \times 10^{-02}$}\\\cline{2-4}
\footnotesize{\textbf{Figure 6a ($E$)}} & \footnotesize{$6.20 \times 10^{+01}$} & \footnotesize{($31.228$, $31$)} & \footnotesize{$4.04 \times 10^{-02}$}\\\hline
\end{tabular}
\caption{Table showing $\mathcal{Z}$ space statistical test results for the semileptonic $t\bar{t}$ dataset. These tests were performed on the distributions in the referenced figures in the main text. ${\rm W}$ is the Wasserstein distance, $\chi_R^2$ is the reduced $\chi^2$ and ${\rm dof}$ is the degrees-of-freedom, and ${\rm KS}$ is the value of the Kolmogorov-Smirnov statistical test. See the Evaluation section in the main text for detailed information about the calculations of these statistics.}
\end{table}

\begin{table}[ht]
\centering
\label{ta:ttbar_xspace}
\begin{tabular}{|p{0.8in}|p{0.8in}|p{0.8in}|p{0.8in}|p{0.8in}|p{0.8in}|p{0.8in}|}
\hline
&\multicolumn{3}{|c|}{$x$ vs $\tilde{x}$}&\multicolumn{3}{|c|}{$x$ vs $\tilde{x}'$}\\\cline{2-7}
& \footnotesize{\textbf{W [GeV$^2$]}} & \footnotesize{\textbf{($\chi^2_R$, dof)}} & \footnotesize{\textbf{KS}} & \footnotesize{\textbf{W [GeV$^2$]}} & \footnotesize{\textbf{($\chi^2_R$, dof)}} & \footnotesize{\textbf{KS}}\\\hline \hline
\footnotesize{\textbf{Figure 6b ($p_y$)}} & \footnotesize{$2.40 \times 10^{+01}$} & \footnotesize{($2.395$, $49$)} & \footnotesize{$1.66 \times 10^{-02}$} & \footnotesize{$1.23 \times 10^{+02}$} & \footnotesize{($34.021$, $49$)} & \footnotesize{$4.59 \times 10^{-02}$}\\\cline{2-7}
\footnotesize{\textbf{Figure 6b ($p_z$)}} & \footnotesize{$1.08 \times 10^{+02}$} & \footnotesize{($0.828$, $55$)} & \footnotesize{$9.90 \times 10^{-03}$} & \footnotesize{$3.42 \times 10^{+02}$} & \footnotesize{($1.980$, $55$)} & \footnotesize{$6.94 \times 10^{-03}$}\\\cline{2-7}
\footnotesize{\textbf{Figure 6b ($E$)}} & \footnotesize{$4.11 \times 10^{+01}$} & \footnotesize{($1.281$, $30$)} & \footnotesize{$1.02 \times 10^{-02}$} & \footnotesize{$3.24 \times 10^{+02}$} & \footnotesize{($50.072$, $30$)} & \footnotesize{$4.80 \times 10^{-02}$}\\\hline
\footnotesize{\textbf{Figure 8a}} & \footnotesize{$1.60 \times 10^{+02}$} & \footnotesize{($1.192$, $43$)} & \footnotesize{$7.63 \times 10^{-03}$} & \footnotesize{$1.03 \times 10^{+03}$} & \footnotesize{($54.598$, $43$)} & \footnotesize{$1.03 \times 10^{-01}$}\\\cline{2-7}
\footnotesize{\textbf{Figure 8b}} & \footnotesize{$9.30 \times 10^{-01}$} & \footnotesize{($3.974$, $35$)} & \footnotesize{$1.66 \times 10^{-02}$} & \footnotesize{$1.04 \times 10^{+02}$} & \footnotesize{($68.392$, $35$)} & \footnotesize{$1.09 \times 10^{-01}$}\\\cline{2-7}
\footnotesize{\textbf{Figure 8c}} & \footnotesize{$8.83 \times 10^{+00}$} & \footnotesize{($1.579$, $30$)} & \footnotesize{$5.91 \times 10^{-03}$} & \footnotesize{$7.41 \times 10^{+01}$} & \footnotesize{($92.533$, $30$)} & \footnotesize{$1.31 \times 10^{-01}$}\\\cline{2-7}
\footnotesize{\textbf{Figure 8d}} & \footnotesize{$2.21 \times 10^{+01}$} & \footnotesize{($2.455$, $41$)} & \footnotesize{$1.72 \times 10^{-02}$} & \footnotesize{$1.11 \times 10^{+03}$} & \footnotesize{($160.712$, $41$)} & \footnotesize{$2.35 \times 10^{-01}$}\\\hline
\end{tabular}
\caption{Table showing $\mathcal{X}$ space statistical test results for semileptonic $t\bar{t}$ dataset. These tests were performed on the distributions in the referenced figures in the main text. ${\rm W}$ is the Wasserstein distance, $\chi_R^2$ is the reduced $\chi^2$ and ${\rm dof}$ is the degrees-of-freedom, and ${\rm KS}$ is the value of the Kolmogorov-Smirnov statistical test. See the Evaluation section in the main text for detailed information about the calculations of these statistics.}
\end{table}

\clearpage
\subsection{Supplementary Schematic Diagram of Network Model}
\begin{figure}[h!]
    \centering
    \begin{subfigure}{0.98\textwidth}
    \includegraphics[width=1.\linewidth]{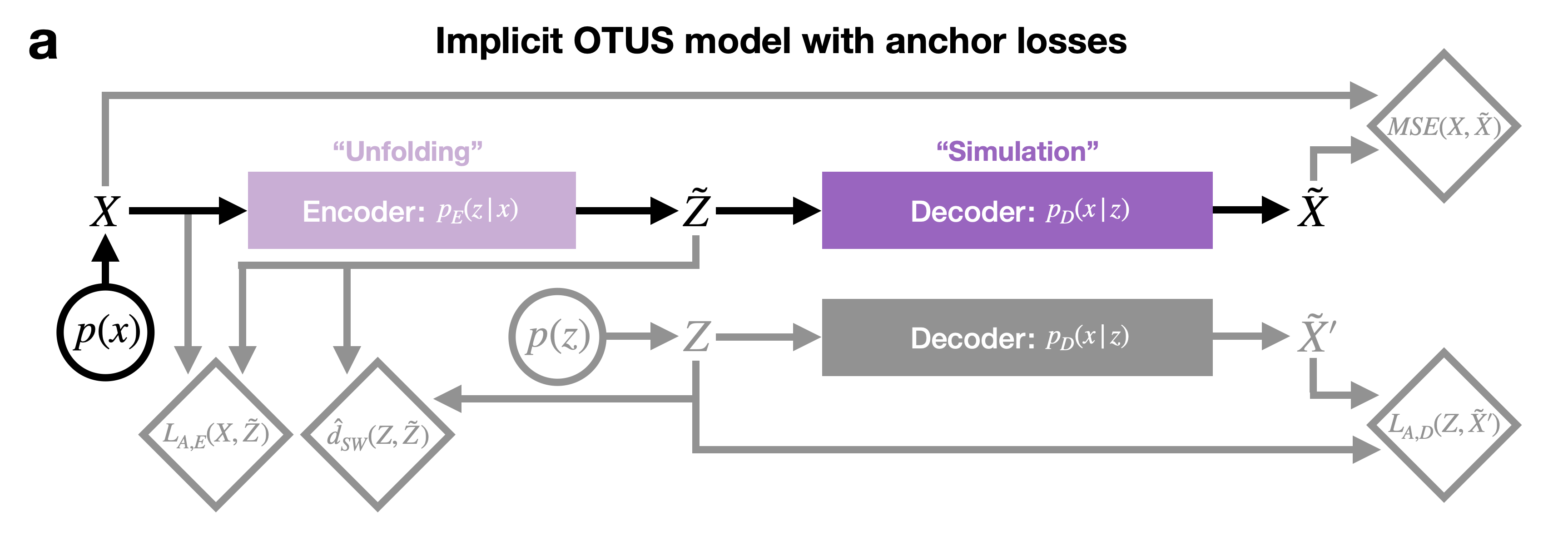}
    \end{subfigure}
    \begin{subfigure}{0.98\textwidth}
    \includegraphics[width=1.\linewidth]{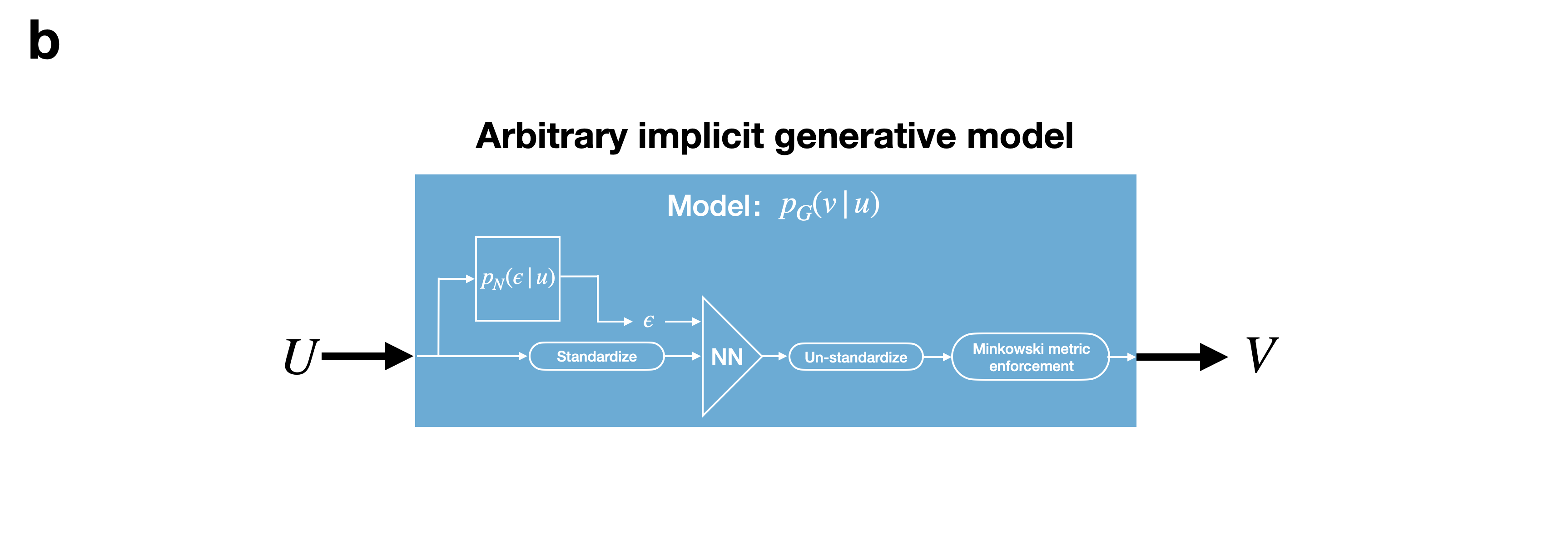}
    \end{subfigure}
    \begin{subfigure}{0.98\textwidth}
    \includegraphics[width=1.\linewidth]{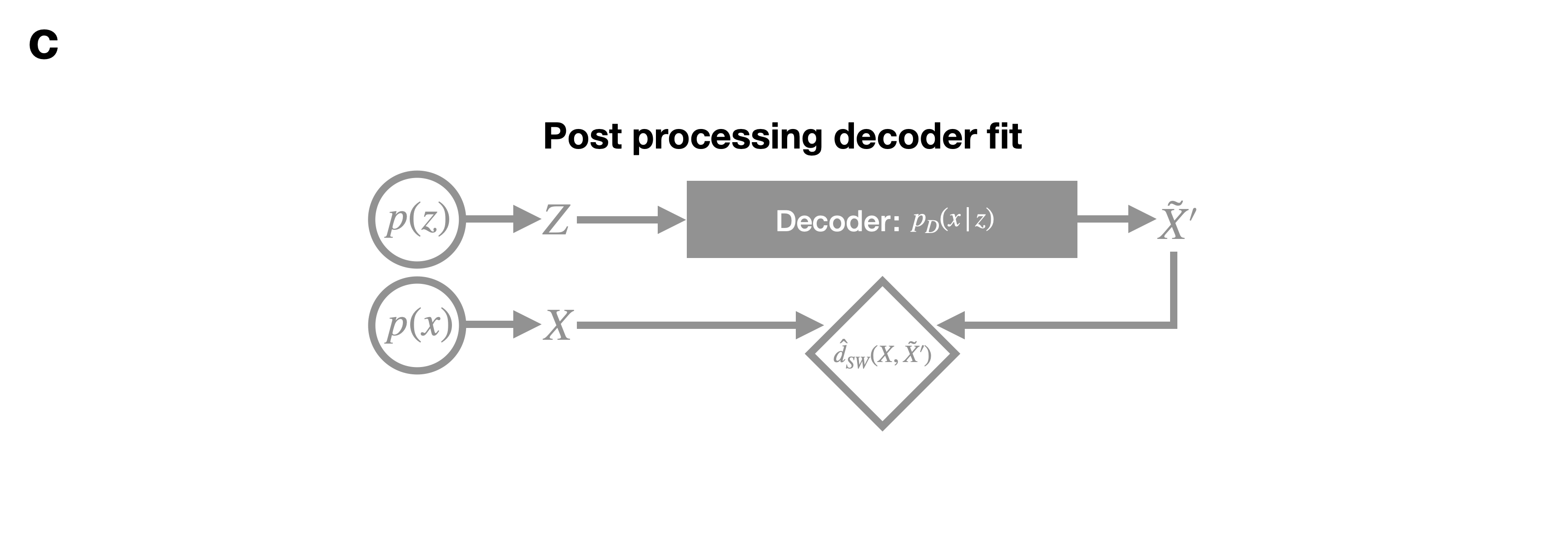}
    \end{subfigure}
    \caption{\textbf{Schematic diagrams of the network and loss structures used in this study for the base training strategy.} 
    \textbf{a} Diagram showing the full OTUS model where gray indicates information used in the calculation of losses only. 
    \textbf{b} Diagram showing the internal structure present in both the encoder and decoder models. 
    \textbf{c} Diagram showing the setup used for the post processing decoder network loss. 
    See the text for more details.}  
    \label{fig:supModel}
\end{figure}

\clearpage
\subsection{Supplementary Ablation Study} \label{subsec:supAblationStudy}

In this section we show the results of an ablation study to demonstrate the effect of the various hyperparameters.
As seen in our final loss function Equation (11),
the main hyperparameters of our approach are the $\lambda$ coefficient in front of the latent space loss, as well as the $\beta_E$ and $\beta_D$ coefficients weighing the anchor losses for the encoder and decoder, respectively. 
For the semileptonic $t \bar{t}$ study the only hyperparameter is $\lambda$, as the anchor loss is redundant with the choice of a ResNet~\cite{ResNet} architecture (see Section 6.2.3). 
We performed ablations by retraining the models as in Section 6.3.3 
but with different values of the hyperparameters on a grid, and comparing the results on validation data. 

For studying the effect of $\lambda$, we reran both the $Z \rightarrow e^+ e^-$ and the semileptonic $t \bar{t}$ studies with $\lambda$ in $\{0.001,$ $0.01,$ $0.1,$ $1,$ $10,$ $100,$ $1000\}$, while keeping all other hyperparamters unchanged (specifically, in the $Z \rightarrow e^+ e^-$ study we kept $\beta_E = \beta_D = 50$). 
For the effect of the anchor loss coefficients, we always assume that $\beta_E = \beta_D$ and define a shared hyperparameter 
$\beta:= \beta_E = \beta_D$. We reran the $Z \rightarrow e^+ e^-$ study with $\beta$ in $\{0,$ $10,$ $20,$ $50,$ $100,$ $200\}$, while keeping $\lambda=1$ as in the original experiment. We did not repeat this for the semileptonic $t\bar{t}$ study as it did not use an anchor loss. 

We first consider how the hyperparameters on the anchor loss terms, $\beta_E = \beta_D$, affect performance. The anchor losses are direct constraints on the learned encoding and decoding mappings which are based on physical concerns. Namely, the anchor loss penalizes networks which would map electron/positron ($e^\mp$) information in $\mathcal{Z}$ to positron/electron ($e^\pm$) information in $\mathcal{X}$, and vice versa. We impose this constraint because we know that misidentification of charge in the process of data reconstruction is extremely rare in particle experiments. Therefore, for our simulation to be physical, it should not make these unphysical inversions. Unsurprisingly, without this constraint we can see that these inversions can occur during training (see Supplementary Figure~\ref{fig:anchorAblation}). On the other hand, if the values of $\beta_E = \beta_D$ are too high we observe unphysical behavior. This is likely due to the fact that the anchor loss is only a proxy for enforcing charge conservation.

\begin{figure}
    \centering
    \includegraphics[width=1.\linewidth]{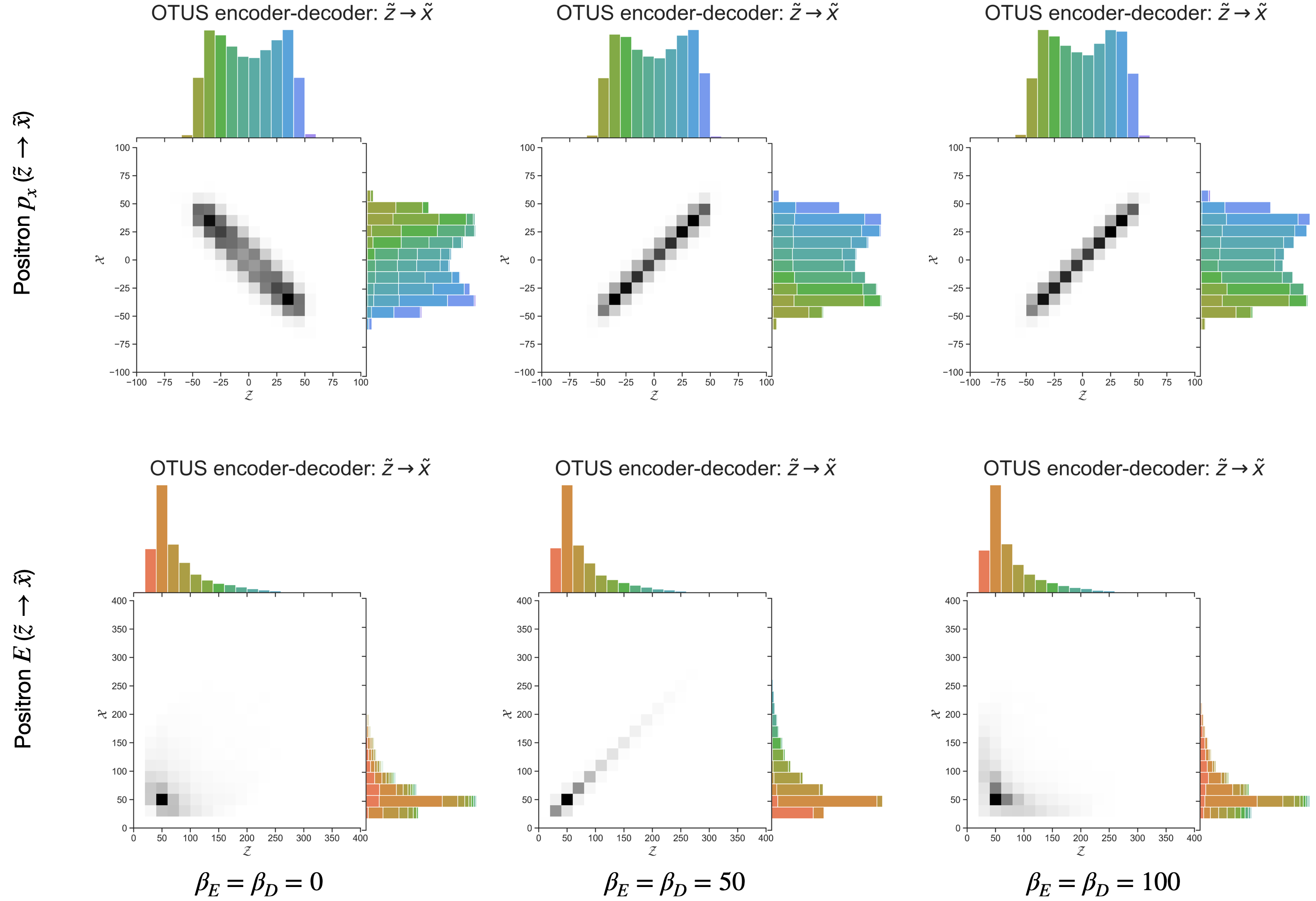}
    \caption{\textbf{Results of anchor loss ablation study in the $Z\rightarrow e^+ e^-$ study.} For $\beta_E = \beta_D = 0$ we can see that unphysical transformations can arise. In $p_x$, negative values in $\mathcal{Z}$ are being mapped to positive values in $\mathcal{X}$. This is a result of $e^\pm$ information being swapped in the learned transformation. For $\beta_E = \beta_D = 50$, this effect goes away; we also see more physical behavior in $E$ as well. For $\beta_E = \beta_D = 100$, we observe that high values of $\beta_E$ and $\beta_D$ inadvertently encourage unphysical behavior in $E$. This is likely due to the fact that the anchor loss is only a proxy for enforcing charge conservation.}  
    \label{fig:anchorAblation}
\end{figure}

We next consider the hyperparameter $\lambda$ which is present in both case studies. The behavior of $\lambda$ has theoretical motivations. 
The WAE method aims to minimize $W_c(p(x), p_D(x))$ by converting its calculation into a constrained optimization problem. It was shown \cite{wae} that $W_c(p(x), p_D(x)) = \inf_{p_E(z|x): p_E(z)=p(z)} \mathbb{E}[c(X, D(Z))]$ for a deterministic decoder $p_D(x|z) = \delta_{D(z)}(x)$%
\footnote{We can show that more generally, for a stochastic decoder, we have an upper bound $W_c(p(x), p_D(x)) \leq \inf_{p_E(z|x): p_E(z)=p(z)} \mathbb{E}_{p(x)p_E(z|x) p_D(\tilde{x}|z)}[c(X, \tilde{X})]$.}. Namely, we need to minimize a reconstruction error over all probabilistic encoders, $p_E(z|x)$, satisfying the latent-space matching condition, $p(z) \stackrel{!}{=} p_E(z) =: \int_x p_E(z|x) p(x) dx$.  To make the constrained optimization computationally tractable, the WAE method only softly enforces this constraint via a penalty term $\lambda d_z(p(z), p_E(z))$, and considers minimizing the surrogate penalty loss $\mathbb{E}_{p(x)p_E(z|x) p_D(\tilde{x}|z)}[c(x, \tilde{x})] + \lambda d_z(p(z), p_E(z))$ instead. 

By standard results on penalty methods~\cite{NocedalAndWright}, for a fixed decoder, $p_D(x|z)$, globally minimizing the penalty loss with respect to the encoder $p_E(z|x)$ results in a lower bound on $W_c(p(x), p_D(x))$, and solving a sequence of such penalized problems while annealing $\lambda$ towards infinity results in the exact $W_c(p(x), p_D(x))$.
However, when training a WAE, it is expensive to repeat this inner optimization procedure after every decoder update, so in practice both the encoder and decoder are optimized jointly on a penalty loss, keeping $\lambda$ fixed throughout the entire training \cite{wae}.

While the theoretical guarantees of the penalty method no longer applies to the joint Stochastic Gradient Descent training procedure used in practice, it does suggest that $\lambda$ should be set to be as large as possible (and perhaps annealed during training) to better enforce the latent space matching, and consequently offer a better approximation of the ideal objective $W_c(p(x), p_D(x))$. Indeed, recently it was proven \cite{sae} that perfect latent space matching $p_E(z) == p(z)$ is a necessary condition for $W(p(x), p_D(x)) = 0$. 

Overall, our ablation experiments confirmed this notion and showed that when $\lambda$ is too small and thus the penalty on latent space matching too week, neither the encoder or the decoder's marginal distribution ($p_E(z), p_D(x)$) could capture the ground truth $p(z)$ or $p(x)$ well, despite minimal reconstruction error.
 
We see this behavior in both test cases, however we note that in the semileptonic $t \bar{t}$ the behavior is somewhat less dramatic because of the heavy initial bias towards an identity mapping due to the ResNet~\cite{ResNet} architecture (see Supplementary Figure~\ref{fig:lambdaAblation_PA}).

\begin{figure}
    \centering
    \includegraphics[width=0.75\linewidth]{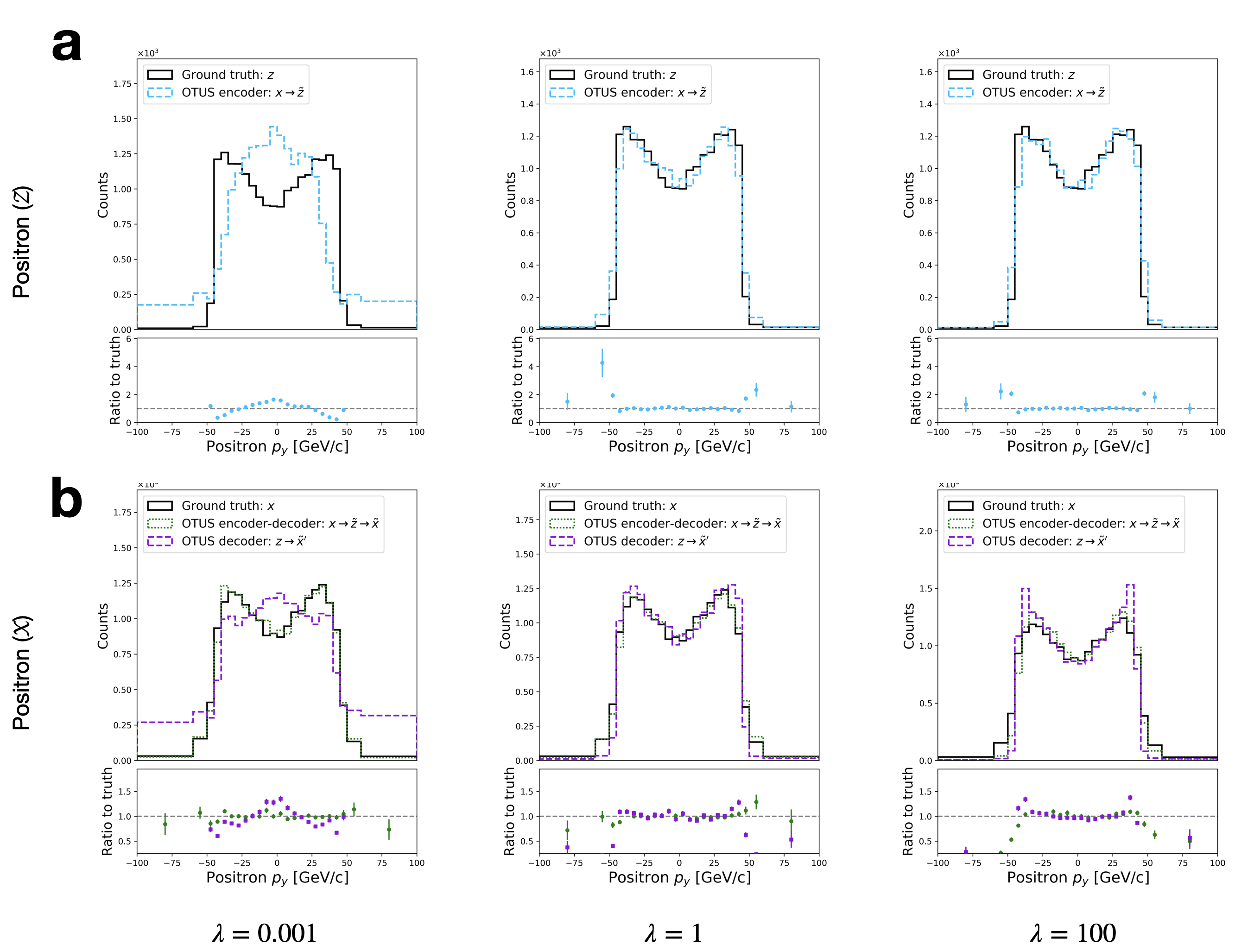}
    \includegraphics[width=0.75\linewidth]{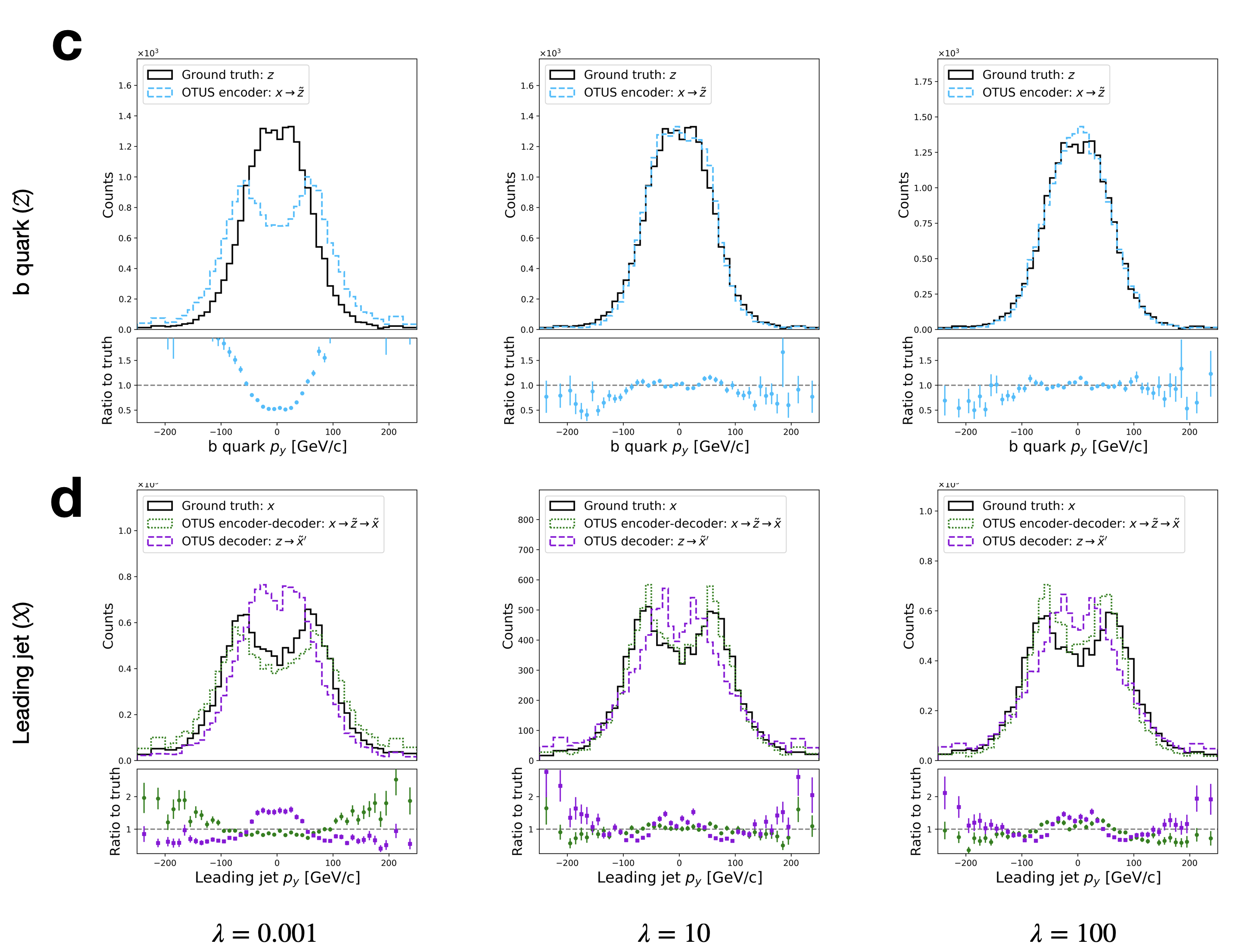}
    \caption{\textbf{Results of $\lambda$ ablation study on principal axis matching.} 
    \textbf{a} Matching of the positron's $p_y$ distribution for $\lambda=0.001$, $\lambda=1$, and $\lambda=100$ in $\mathcal{Z}$ for the $Z \rightarrow e^+ e^-$ study.
    \textbf{b} Matching of the positron's $p_y$ distribution for $\lambda=0.001$, $\lambda=1$, and $\lambda=100$ in $\mathcal{X}$ for the $Z \rightarrow e^+ e^-$ study.
    \textbf{c} Matching of the $b$ quark's $p_y$ distribution for $\lambda=0.001$, $\lambda=10$, and $\lambda=100$ in $\mathcal{Z}$ for the semileptonic $t\bar{t}$ study.
    \textbf{d} Matching of the leading jet's $p_y$ distribution for $\lambda=0.001$, $\lambda=10$, and $\lambda=100$ in $\mathcal{X}$ for the semileptonic $t\bar{t}$ study.
    For small values of $\lambda$ ($\lambda=0.001$) we find that performance suffers as latent space matching is not enforced. This improves as we increase $\lambda$ but eventually plateaus.
    }  
    \label{fig:lambdaAblation_PA}
\end{figure}

\begin{figure}
    \centering
    \includegraphics[width=0.85\linewidth]{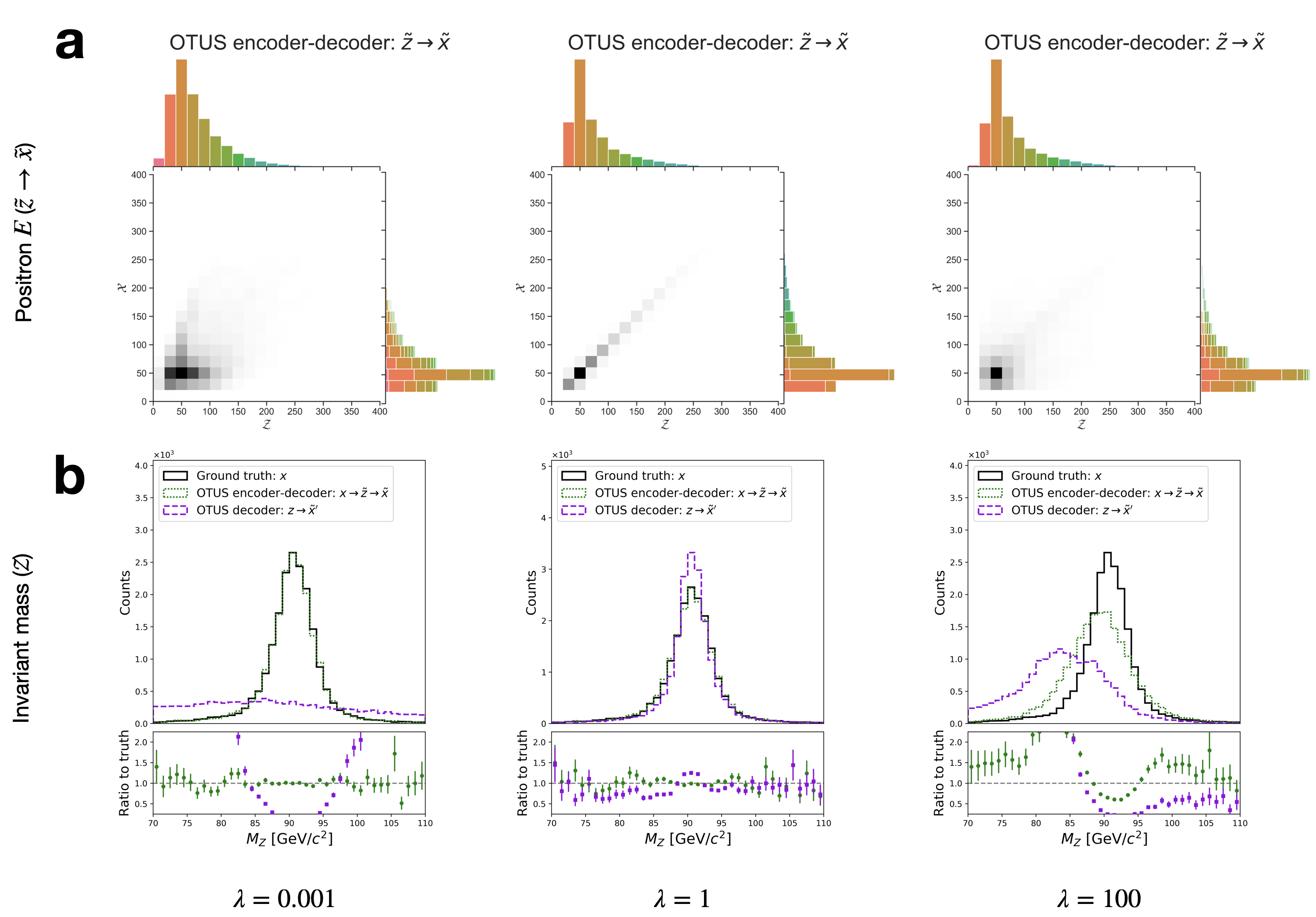}
    \includegraphics[width=0.85\linewidth]{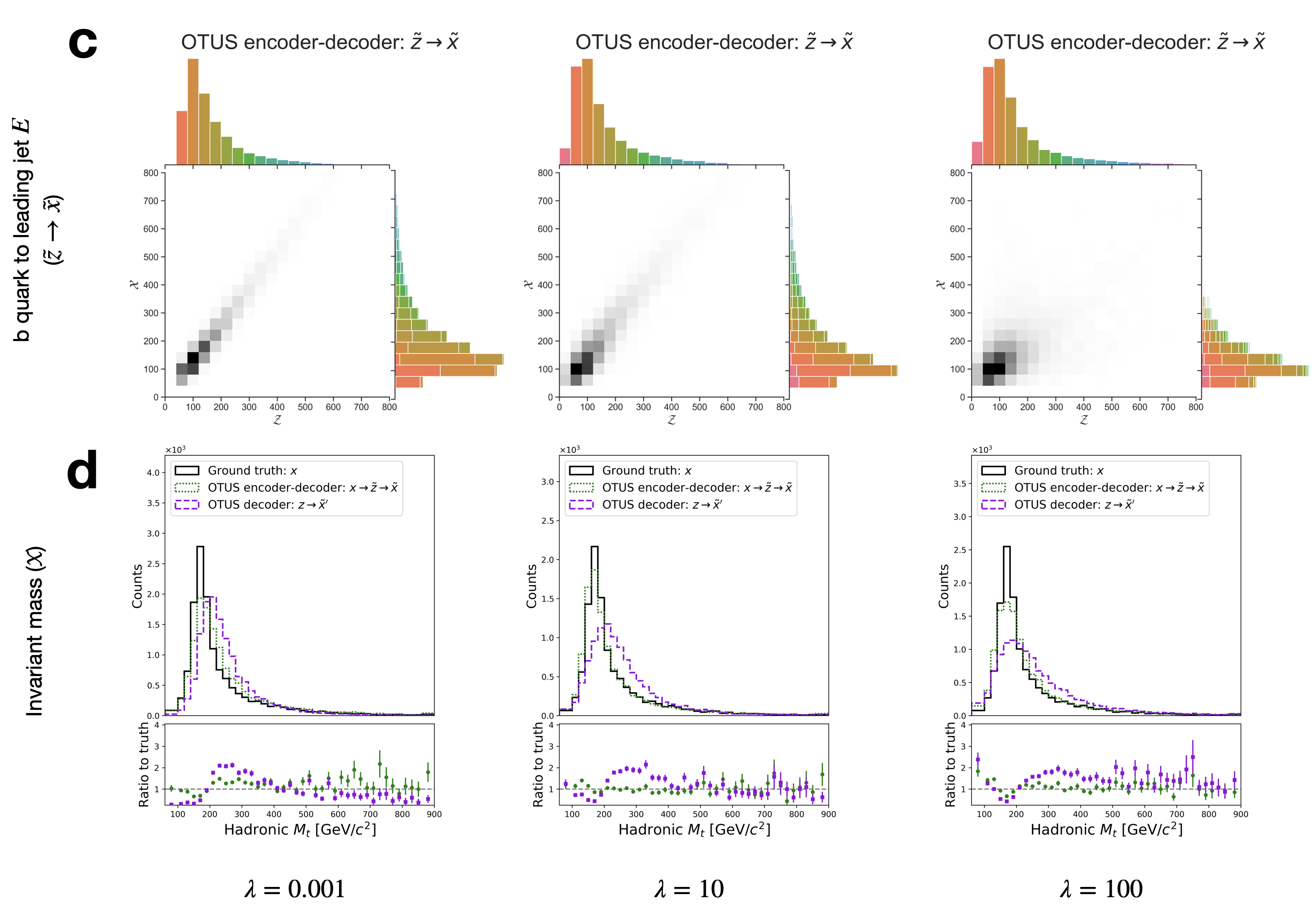}
    \caption{\textbf{Results of $\lambda$ ablation study on transport plots and derived quantity matching.}
    \textbf{a} Transport plans from $\tilde{z} \rightarrow \tilde{x}$ of the positron's $E$ distribution for $\lambda=0.001$, $\lambda=1$, and $\lambda=100$ for the $Z \rightarrow e^+ e^-$ study.
    \textbf{b} Matching of the invariant mass of the $Z$-boson for $\lambda=0.001$, $\lambda=1$, and $\lambda=100$ for the $Z \rightarrow e^+ e^-$ study.
    \textbf{c} Transport plans from $\tilde{z} \rightarrow \tilde{x}$ of the b quark's $E$ distribution in $\mathcal{Z}$ to the leading jet's $E$ distribution in $\mathcal{X}$ for $\lambda=0.001$, $\lambda=10$, and $\lambda=100$ for the semileptonic $t\bar{t}$ study.
    \textbf{d} Matching of the invariant mass of the top-quark, $M_t$, reconstructed using information from the hadronically decaying $W$-boson for $\lambda=0.001$, $\lambda=10$, and $\lambda=100$ for the semileptonic $t\bar{t}$ study.
    }  
    \label{fig:lambdaAblation_derived}
\end{figure}
We see performance in matching principal axes improve as $\lambda$ grows larger, possibly plateauing in the case of the semileptonic $t \bar{t}$ study. This plateau is potentially due to issues with optimization and poor numerical conditioning with overly large $\lambda$. However, we find that too large of a value of $\lambda$ results in unphysical mappings.

Specifically, we find unphysical behavior when we view the transport plots and derived quantities (see Fig~\ref{fig:lambdaAblation_derived}). Again, we note that this is less noticeable for the semileptonic $t \bar{t}$ study due to the ResNet~\cite{ResNet} architecture. We find that the ideal choice is $\lambda \approx 1$ for the $Z \rightarrow e^+ e^-$ study and $\lambda \approx 20$ for the semileptonic $t\bar{t}$ study; this retains acceptable principal axis matching while not introducing unphysical transformation characteristics. We suspect that if the choice of $\lambda$ is too large, it over-constrains the optimization problem and should instead be annealed.

As discussed, instead of an expensive double-loop procedure where we train the encoder to optimality with the penalty method before updating the decoder,  we forego theoretical considerations by jointly optimizing the encoder and decoder of a WAE on a surrogate loss as in \cite{wae}. The resulting loss is neither an upper nor a lower bound on the ideal objective $W_c(p(x), p_D(x))$, and we choose $\lambda$ by experimentation.
An alternative would be to use the Sinkhorn Autoencoder~\cite{sae} approach, which only needs a large enough $\lambda$ for its loss to be a proper upper bound on $W_c(p(x), p_D(x))$. This is further motivation that this method should be explored in future work.


\typeout{get arXiv to do 4 passes: Label(s) may have changed. Rerun}
\end{document}